\pdfoutput=1
\documentclass{JINST}
\usepackage{capt-of}

\title{Operation and performance of the ICARUS-T600 cryogenic plant at Gran Sasso underground Laboratory}

\author{
M.~Antonello$^a$,
P.~Aprili$^a$,
B.~Baibussinov$^b$,
F.~Boffelli$^c$,
A.~Bubak$^d$,            
E.~Calligarich$^c$,
N.~Canci$^a$,
S.~Centro$^b$,
A.~Cesana$^e$,           
K.~Cie\'slik$^f$,        
D.B.~Cline$^g$,          
A.G.~Cocco$^h$,          
A.~Dabrowska$^f$,        
A.~Dermenev$^i$,         
J.M.~Disdier$^p$,
A.~Falcone$^c$,
C.~Farnese$^b$,
A.~Fava$^b$,
A.~Ferrari$^j$,              
D.~Gibin$^b$,
S.~Gninenko$^i$,         
A.~Guglielmi$^b$,
M.~Haranczyk$^f$,        
J.~Holeczek$^d$,         
A.~Ivashkin$^i$,         
M.~Kirsanov$^i$,         
J.~Kisiel$^d$,           
I.~Kochanek$^d$,         
J.~Lagoda$^k$,           
S.~Mania$^d$,            
A.~Menegolli$^c$,
G.~Meng$^b$,
C.~Montanari$^c$,
S.~Otwinowski$^g$,      
P.~Picchi$^l$,
F.~Pietropaolo$^b$,
P.~Plonski$^m$,
A.~Rappoldi$^c$,
G.~L.~Raselli$^c$,
M.~Rossella$^c$,
C.~Rubbia$^{a,j}$,       
P.~R.~Sala$^n$,
A.~Scaramelli$^n$,
E.~Segreto$^a$,
F.~Sergiampietri$^o$,
D.~Stefan$^n$,
R.~Sulej$^k$,            
M.~Szarska$^f$,          
M.~Terrani$^e$,          
M.~Torti$^c$,
F.~Varanini$^b$,
S.~Ventura$^b$,
C.~Vignoli$^{a,*}$,
H.G.~Wang$^g$,           
X.~Yang$^g$,             
A.~Zalewska$^f$,         
A.~Zani$^c$,
K.~Zaremba$^m$\\
\llap{$^a$}INFN - Laboratori Nazionali del Gran Sasso, Assergi, Italy\\
\llap{$^b$}Universit\`a di Padova e INFN,  Padova, Italy\\
\llap{$^c$}Universit\`a di Pavia e INFN, Pavia, Italy\\
\llap{$^d$}Institute of Physics, University of Silesia, Katowice, Poland\\
\llap{$^e$}INFN e Politecnico di Milano, Milano, Italy\\
\llap{$^f$}H.Niewodnicza\'nski Institute of Nuclear Physics, Krak\'ow, Poland\\
\llap{$^g$}Department of Physics, UCLA, Los Angeles, USA\\
\llap{$^h$}Universit\`a Federico II di Napoli e INFN, Napoli,\\
\llap{$^i$}Institute for Nuclear Research of the Russian Academy of Sciences, Moscow, Russia\\
\llap{$^j$}CERN, Geneva, Switzerland\\
\llap{$^k$}National Centre for Nuclear Research, Otwock, \'Swierk, Poland\\
\llap{$^l$}INFN Laboratori Nazionali di Frascati, Frascati, Italy\\
\llap{$^m$}Institute for Radioelectronics, Warsaw Univ. of Technology, Warsaw, Poland\\
\llap{$^n$}INFN Sezione di Milano, Milano, Italy\\
\llap{$^o$}Universit\`a di Pisa e INFN, Pisa, Italy\\
\llap{$^p$}Luca Scarcia Company, Italy\\
\llap{$^*$}Corresponding author; e-mail: \email{chiara.vignoli@lngs.infn.it}}
\abstract{ICARUS T600 liquid argon time projection chamber is the first large mass electronic detector of a new  generation able to combine the imaging capabilities of the old bubble chambers with the excellent calorimetric energy measurement.  After the three months demonstration run on surface in Pavia during 2001, the T600 cryogenic plant was significantly revised, in terms of reliability and safety, in view of its long-term operation in an underground environment. The T600 detector was activated in Hall B of the INFN Gran Sasso Laboratory during spring 2010, where it was operated without interruption for about three years, taking data exposed to the CERN to Gran Sasso long baseline neutrino beam and cosmic rays. In this paper the T600 cryogenic plant is described in detail together with the commissioning procedures that lead to the successful operation of the detector shortly after the end of the filling with liquid Argon.Overall plant performance and stability during the long-term underground operation are discussed. Finally, the decommissioning procedures, carried out about six months after the end of the CNGS neutrino beam operation, are reported.}
\begin{document}

\section{Introduction}

The ICARUS T600 liquid argon time projection chamber (LAr-TPC)~\cite{t600_jinst} is the largest LAr
 imaging detector ever built with a total argon mass of $\sim 760$ t. It was located at the INFN Gran Sasso 
 underground Laboratory (LNGS) with a coverage of 1400 meters of rock. The operational principle of 
 the LAr-TPC~\cite{rubbia77} is based on the possibility, in highly purified LAr, 
 to transport free electrons from ionizing tracks practically undistorted by a uniform electric field over macroscopic distances. 
 A suitable set of electrodes (wires) placed at the end of the drift path continuously sense and record the signals induced by 
 the drifting electrons. This provides  simultaneous projective views of the same event, allowing precise three-dimensional 
 imaging capability~\cite{3d} and high resolution calorimetric measurements. The design and assembly of the ICARUS-T600 LAr-TPC relied 
 on industrial support and represents the application of concepts matured in laboratory tests to the kton scale.

 The T600 was smoothly and safely operated from May 2010 to June 2013, taking data on the CERN to Gran Sasso (CNGS)
 neutrino beam with extremely high argon purity, stability and detector live-time~\cite{t600_jinst}. It also acted 
 as underground observatory  recording cosmic ray and atmospheric neutrino events.
The T600 detector has a unique role since it is presently the largest physics grade operational LAr-TPC and it 
 will remain so for several years to come. It represents the state of the art and it marks a milestone in the 
 practical realisation of any future larger scale LAr detector. 
 The successful operation of the ICARUS T600 LAr-TPC, which allowed
to perform a sensitive search for  $\nu_\mu \rightarrow \nu_e$ oscillations~\cite{numu-nue2,numu-nue1}, 
 demonstrates the enormous potential of this detection technique.
 The procedures that brought to the activation and operation of the T600 cryogenic plant in a difficult underground 
 environment open the way to  larger detector masses (up to tens of ktons) as foreseen by several new neutrino and 
 rare event physics projects.
 
 The ICARUS T600 detector  was moved to CERN in 2014 for overhauling. It is expected to be put again in operation at 
 FNAL exposed to the short Booster neutrino beam~\cite{CERNProposals}, for a definitive
 clarification of  the observed  neutrino anomalies~\cite{LSNDres, Minibooneres, numu-nue2} hinting at 
 the presence of a new ``sterile'' neutrino state. It will also collect a large sample ($\geq 10^6$) of neutrino interactions 
 from the NuMI beam in the few GeV energy domain relevant to the future long base-line experiment~\cite{lbneexp}, allowing detailed study of any event 
 topology and precise tuning of the reconstruction tools.
 The foreseen LAr program may also pave the way to ultimate realization of the multi kton detector with the precise
 measurement of the visible energies of both hadron and electron showers and of the muon momentum determination~\cite{lbneexp, 
 modular1, glacier, icanessie, microboone}. 
 
In this paper the T600 cryogenic plant is described in details together with all the underground infrastructures, 
as well as the main phases of ICARUS T600 commissioning, which allowed collecting cosmic ray and CNGS neutrino 
events soon after the detector filling with ultra-pure LAr. Steady state operation is reported focusing on the
stability and reliability of the plant. The decommissioning phase is also described.

\section{Overview of the ICARUS T600 liquid argon TPC}
\label{overview}
The ICARUS T600 plant~\cite{t600}  consists of a large cryostat split into two identical, 
adjacent ``modules'' with internal dimensions 3.6 (width) $\times$ 3.9 (height) $\times$ 19.6 (length) m$^3$ and 
filled with $\sim 760 $ tons of ultra-pure liquid argon. 

Each module houses two TPC's separated by a common cathode, with a drift length of 1.5 m. 
Ionization electrons, abundantly produced by charged particles along their path ($\sim$ 5000 electrons/mm for 
minimum ionizing particles), are drifted under uniform electric field (E$_D$ = 500 V/cm) towards the TPC anode made 
of three parallel wire planes, 3 mm apart, facing the drift volume. A total of 53248 wires are deployed, with 3 mm pitch, 
oriented on each plane at a different angle (0$^{\circ}$, +60$^{\circ}$, -60$^{\circ}$) with respect to the horizontal direction. 
Wires are made of AISI 304V stainless steel with a diameter of 150 ${\mu}$m and maximum length of 9.42 m for the 
horizontal wires or 3.77 m for the inclined ones. The wire-frame mechanics is based on the innovative concept of variable geometry 
design consisting in movable and spring-loaded frames to set the proper tension of the wires after installation for precise detector geometry 
and planarity, to compensate for possible over-stress during the cooling-down and liquid argon filling phases and 
to counteract the flexibility of the frame.  This design  demonstrated its reliability, since none of the wires 
broke and no damages at the wire chamber structure occurred during the 2001 test run, the transport of the two 
modules from Pavia INFN Laboratory to LNGS, the installation movements on site, the commissioning phase at LNGS 
and all the successive operations.

By appropriate voltage biasing, the first two wire planes (``Induction-1'' and ``Induction-2'') provide signals 
in non-destructive way; finally the ionization charge is collected and measured on the last plane (``Collection''). 
The relative time of each ionization signal, combined with the electron drift velocity information 
(v$_D$ $\sim$ 1.6 mm/$\mu$s), provides the position of the track along the drift coordinate. 
Combining the wire coordinate on each plane at a given drift time, a three-dimensional image of 
an ionizing event can be reconstructed with the remarkable resolution of about 1 mm$^3$. 

A special feed-through flange\footnote{Proprietary technology from INFN.} for the wire signals has been 
adopted in the T600 detector. It is based on multilayer printed circuit boards where the electrical
 contacts are ensured by blind holes realized staggering successive printed circuit boards (PCB) layers. 
 The absence of through-going holes ensures perfect tightness for Ultra High Vacuum (UHV) applications. 
 This design was successfully tested in the WARP experiment~\cite{warp} at LNGS showing high reliability.  
 A set of 96 flanges, holding 576 channels each,  was installed on the T600 during detector reassembly at LNGS.

The read-out electronic chain was designed to allow continuous read-out, digitization and independent waveform 
recording of signals from each wire of the TPC. Organized in 96 crates placed on top of the detector
cryostat, it provided wire biasing, hosted the front-end amplifiers and performed 16:1 channel multiplexing 
and 10-bit ADC digitization at 400 ns sampling time per channel.  The  electronic noise achieved with the
 custom designed low noise front-end was $\sim$1500 electrons r.m.s. to be compared with $\sim$15000 free 
 electrons produced by minimum ionizing particles in the 3 mm wire pitch (S/N $\sim$10).

The absolute time of the ionizing event was provided by the prompt VUV (128 nm) scintillation light emitted 
in liquid argon ($\sim$ 5000 photons/mm for minimum ionizing particles) and measured through arrays of Photo 
Multiplier Tubes (PMTs) coated with VUV sensitive wavelength-shifter (Tetra-Phenyl-Butadiene, TPB), 
placed in LAr behind the wire planes. The PMT system, used also for internal trigger purposes \cite{trigger-LNGS}, 
allowed for a precise measure of neutrino velocity with CNGS beam~\cite {velocita-nu,velocita-nu0}.

Liquid argon purity is a key issue for the LAr-TPC. LAr electronegative impurities (mainly H$_2$O, O$_2$, CO$_2$)
must be kept at a very low concentration level (less than 0.1 ppb), to allow ``unperturbed'' drift of ionization 
electrons  from the point of production to the wire planes. To this purpose the ICARUS Collaboration developed a
successful technique based on the use of commercial filters, 
carefully selected among the many availabilities on the market and operated directly on liquid and
the adoption of ultra high vacuum materials and techniques. The commissioning procedures included an initial 
vacuum phase followed at regime by the continuous argon re-circulation and purification of both the liquid bulk
and the top gas phase.  

Additional details on the T600 detector design and construction can be found in~\cite{t600}; initial detector 
operation and performance  at LNGS exposed to CNGS neutrino beam and cosmic rays are described in~\cite{t600_jinst}.

\subsection{The T600 underground location at LNGS}

\begin{figure}[htbp]
\centering
\includegraphics[width=14cm]{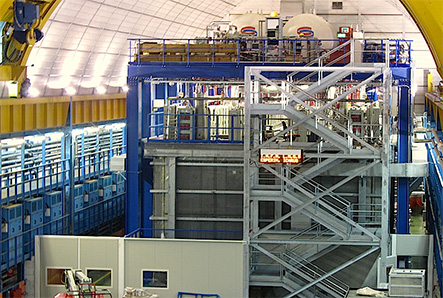}
\caption{The ICARUS T600 plant in the Northern side of the Hall B at LNGS. Ladders to access the intermediate 
and the top levels are visible in the front together with the 3 m high wall that separates the T600 area from 
the rest of Hall B. Cabinets with the readout electronics are visible at the intermediate level. 
At the top level, the two 30 m$^3$ cryogenic storage tanks and some additional electronic equipments (HV supply, 
PMT electronics, trigger system, etc.) are also visible. Most of the cryogenic equipment (nitrogen and argon pumps, 
cryo-coolers, etc.) is located on the rear side of the T600, at the Northern end of the Hall B.}
\label{icarus}
\end{figure}

\begin{figure}[htbp]
\centering
\includegraphics[width=15cm]{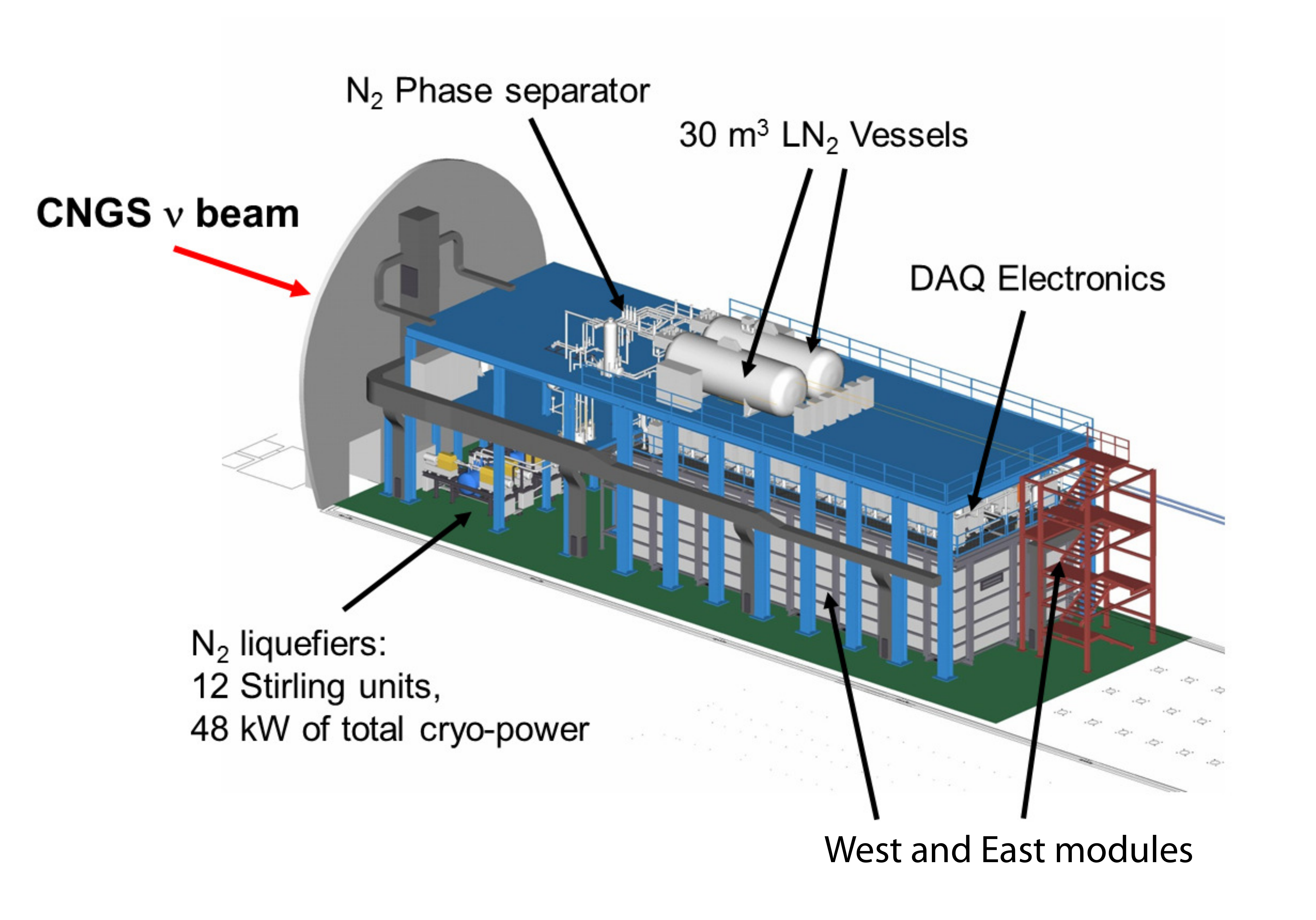}
\caption{Schematic view of the whole ICARUS T600 plant in the Hall B at LNGS.}
\label{icasketch}
\end{figure}

The ICARUS T600 plant and its dedicated technical infrastructures were installed in the Northern-end side of the
Hall B of the underground LNGS Laboratory (Fig.~\ref{icarus}, Fig.~\ref{icasketch}). The final design of the apparatus 
was strongly affected by strict requirements on efficiency, safety, 
seismic constraints and reliability for long time operation in a confined underground experimental 
area at 1400 m depth. Further severe prescriptions arose from the specific position of the Laboratory 
along a 10.5 km long highway tunnel with the access in the middle of the tunnel. Additional restrictions 
came from the LNGS Laboratory location inside a National Park and from the proximity of a public aqueduct.

The T600 plant occupied about 10 $\times$ 22 m$^2$ surface, while the whole ``ICARUS Area'' in the Hall B 
was 15  $\times$ 42,5 m$^2$.  This area contained a dedicated service structure surrounding the T600 
(about 11 $\times$ 36  m$^2$) organized in three levels (floor level, 5 and 10 m height) in order to maximize the 
use of available space in the semi-cylindrical hall. It hosted two 30 m$^3$ horizontal  cryogenic liquid 
storages (top level), the nitrogen re-liquefaction apparatus (rear floor) and most of the ICARUS auxiliary 
systems. The ICARUS Area was served by the two Hall B cranes (40 t and 5 t). 
The apparatus was placed on 28 dampers to fulfill the seismic requirements of the Gran Sasso region. 
As a reference,  the 5.8 Richter magnitude earthquake occurred in the L'Aquila region in 2009 did not produce 
any appreciable effects on the T600 plant. The ICARUS Area, adequately equipped with a fine grid of sensors 
to promptly detect any presence of liquids, 
oxygen deficiency and temperature decrease, was separated from the rest of the Hall B by means of 
a 3 m high wall for safety reasons. A capillary fast extraction aspiration 
from ground to protect plant and personnel in case of cryogenic liquid and cold gas spillages, as well as many 
other safety sensors such as fire and smoke monitors were also installed.
The ICARUS Control Room, which hosted all the data acquisition systems and the remote controls of the cryogenic 
plant and of the detector, was located on the Southern side of the Hall B.

Most of the infrastructures and auxiliary systems were specifically developed for the T600 operation at LNGS, 
such as redundant power supply sources and distribution systems, uninterrupted power supply for cryogenic plant 
and detector control systems, upgraded and redundant cooling water system and the nitrogen re-liquefaction plant.

\subsection{The T600 cryogenic plant}
\label{T600cryo}

To achieve the physics goals of the ICARUS experiment, the T600 design had to fulfill several strict 
requirements in terms of detector mechanics precision and stability, electronic noise, argon purity 
and cryogenic plant performance and reliability. The main technical specifications for the T600 cryogenic plant were:

\begin{itemize}

\item[-] an extremely high liquid argon purity in terms of residual contamination of electronegative molecules such as water,
 oxygen and carbon dioxide (better than 0.1 part per billion) to allow ionization electrons to drift over long distances (meters);

\item[-] a fast cooling from room to liquid Ar temperature to minimize outgassing and obtain a good initial liquid 
argon purity while guaranteeing a temperature difference within each detector component  within prescribed limits 
($\Delta$T $<$ 50 K on the wire-chamber structures, $\Delta$T $<$ 120 K on the cold vessels)
to preserve the high precision detector mechanics;

\item[-] a very high temperature uniformity in steady state conditions ($\Delta$T $<$ 1 K in the whole liquid argon volume) 
to guarantee the uniformity of the electron drift velocity;

\item[-] low and stable thermal losses to reduce operation costs and minimize the power request in emergency situations;

\item[-] a very high stability and operation reliability to fulfill the strict underground safety requirements;

\item[-] a full redundancy of the relevant elements to guarantee uninterrupted operation over several years;

\item[-] no electronic noise or vibration introduced on detector by cryogenic plant dynamic components operation;

\item[-] a redundant cryogenic containment for safety aspects;

\item[-] an availability of spares components to garantee long term operation.

\end{itemize}

The common denominator of the overall ICARUS T600 construction was the partnership with industry based 
on the achievements and technical developments from the long ICARUS R\&D experience with different 
prototypes~\cite{3ton1,3ton2,10m3} and the use of commercial solutions available on the market, 
when possible. Industry involvement was crucial in order to perform a scaling-up of the technology from the 
laboratory prototypal scale to the kton one. In particular the T600 cryostat design and construction was carried out in strict 
collaboration with Air Liquide Italia Service (ALIS) Company\footnote{www.airliquide.it}.

\begin{figure}[htbp]
\centering
\includegraphics[width=13cm]{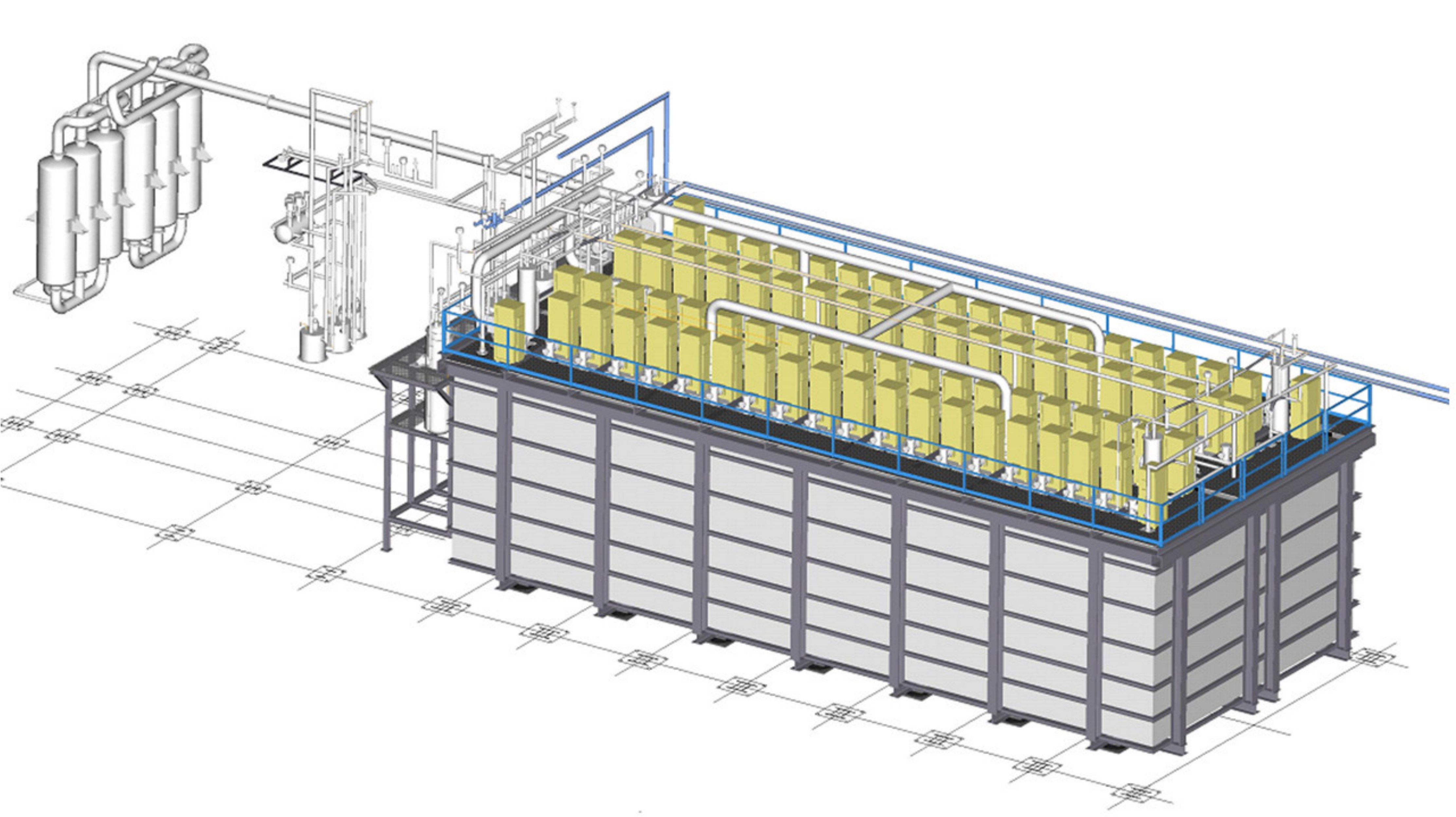}
\caption{Schematic view of the T600 cryostat with representation of the electronic racks and gaseous argon 
re-circulation systems on the top, the liquid nitrogen  circulation pumps and liquid argon re-circulation 
systems in the rear side. The major safety equipments are visible: the vent line with the electrical heater 
connected to all the possible exhausts and the collecting pipes from the T600 modules and insulation vessel 
safety disks to the  six passive heater system.}
\label{t6003d}
\end{figure}

Safety requirements for underground run implied major changes and improvements with respect to the operational 
conditions carried out during the surface test in the Pavia INFN Laboratory in 2001~\cite{t600}. In particular new 
solutions for cooling and insulation systems were adopted~\cite{vignoli}.
The final T600 design at LNGS was further conditioned by the choice of building major components outside the tunnel 
to work in a more effective and comfortable way and to reduce interferences with other 
activities and experiments running underground. Schematic views of the T600 plant at LNGS are shown in 
Fig.~\ref{icasketch} and Fig.~\ref{t6003d}.  
In the rest of the paper we describe in details only the relevant innovations of the cryogenic plant while for the 
unchanged components we refer to ~\cite{t600}.

\begin{figure}[htbp]
\centering
\includegraphics[width=15cm]{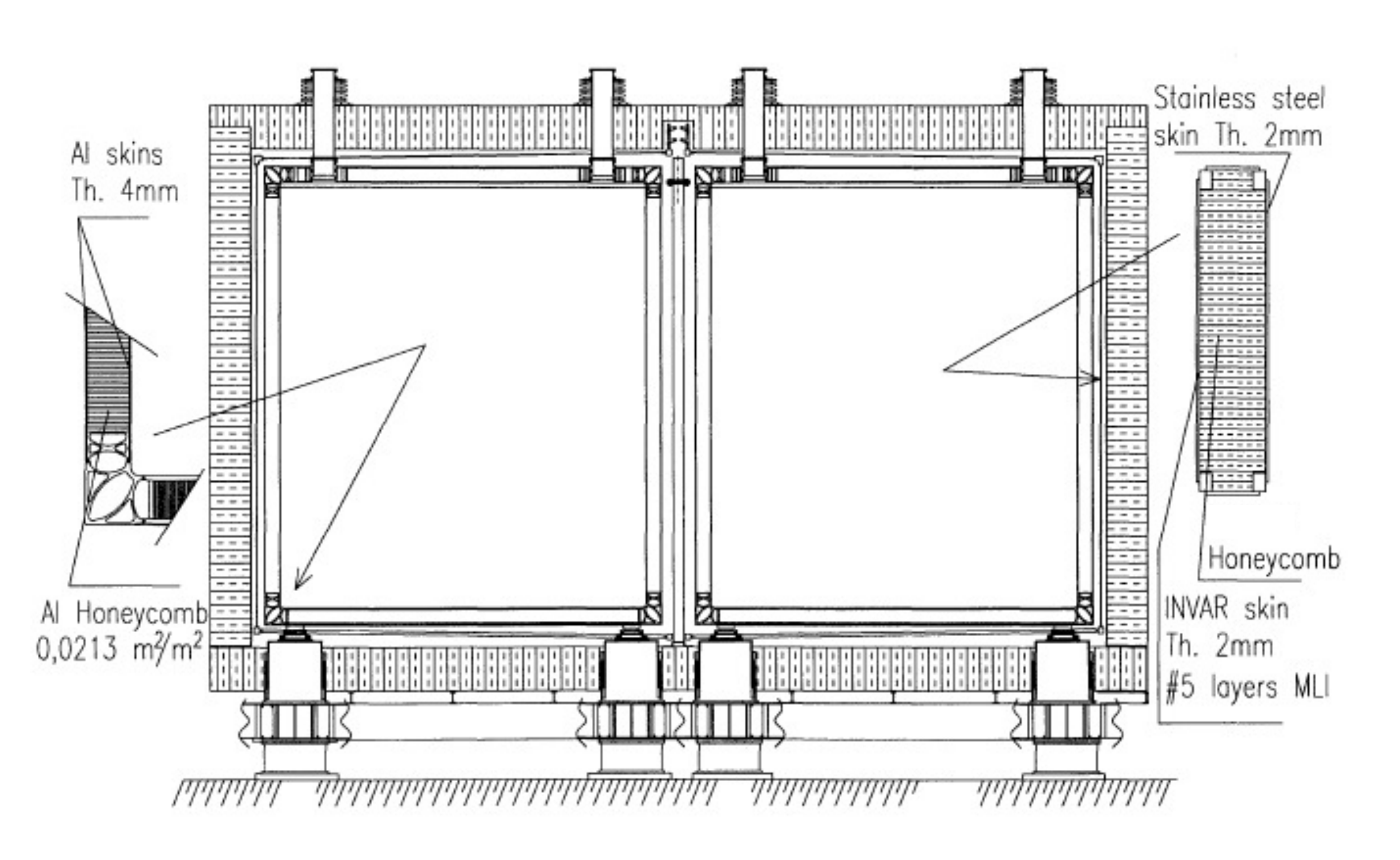}
\caption{\label{sketch} Vertical cross section of the ICARUS T600 cryostat.}
\end{figure}

The two T600 modules were independent from the point of view of LAr containment and purification plants (Fig.~\ref{sketch})
while the nitrogen cooling system and the thermal insulation were common to both. 
The two modules were aluminum parallelepiped containers, each one with an internal volume of 275 m$^3$ and 4.2 h 
$\times$ 3.9 w $\times$ 19.9  l m$^3$ external dimensions, had the
maximum size allowed for transportation into the LNGS underground laboratory.  In the following they 
will be referred as West module - the one already commissioned in 2001 in Pavia and widely characterized~\cite{pi0-pavia, ms-pavia, birks-pavia, michel-pavia, purity-pavia, traccione-pavia} - and East module (the CNGS 
beam was arriving from the North, Fig.~\ref{icasketch}). 
Each cold vessels was realized with 15 mm thick aluminum honeycomb panels linked to Al skins 
and to extruded profiles at the borders with the external and internal skins acting as double cryogenic containment for safety.
 This unconventional vessel solution was adopted mainly for lightness request for transportability, rigidity to stand stresses during the evacuation phase and the overall LAr and detector weight during steady state ~\cite{t600}. 

The T600 thermal insulation was a single vessel surrounding the two modules and designed to behave as an additional 
container  for safety in case of cryogenic liquid spillages. Insulation vessel walls, with the exception of the roof,
were made of metallic boxes filled with insulating honeycomb panels (0.4 m thick of Nomex$^{TM}$ or equivalent material)
and super-insulation layers placed on the internal (cold) surface \cite{vignoli}.
The outer skins of the boxes were made of stainless steel while the inner and side skins were of Pernifer$^{TM}$  to 
avoid thermal shrinking, Each compartment was designed to be operated in vacuum  to reduce gas conduction
and convection (limited by the honeycomb cell geometry but still present due to residual gas) while radiative losses
were suppressed by the super-insulation layers, resulting in an overall heat load around 10 W/m$^2$. 
They were intended to be evacuated only once down to 10$^{-4}$ mbar and then kept in static vacuum conditions by 
means of getter pumps. 
However various difficulties introduced by the large wall dimensions  together  with an excessive outgassing of the 
internal honeycomb surface (mainly water) were faced during the construction and mounting of the insulation vessel 
 preventing to reach the nominal design 
 parameters\footnote{A mechanical instability problem occurred during evacuation of the Northern insulation wall,
  causing some domino effects. The North wall was repaired and its internal honeycomb replaced on-site with four 
  closed-cell  Divinycell$^{TM}$ 0.1 m layers. After this event the North and East insulation walls were left at about ambient pressure.}. 
As a consequence,  during cryostat operation, insulation walls were maintained under dynamic evacuation and their 
internal pressure was continuously monitored for safety reasons. 

Holes for pipes and chimneys for cables of the detector signals were all located on the ceiling of the insulation 
vessel matching the 
position of all feed-throughs on the two module ports. Special bellows mounted around all the crossing tubes 
ensured the proper tightness of the box. An external metallic cage placed on an anti-seismic shock-absorber system 
reinforced the 
insulation box guaranteeing stress distribution in case of internal overpressure and proper performance for residual 
accelerations 
during earthquake events. It supported all the weight of the  electronic racks present on the T600 top.

A cooling shield for nitrogen circulation was placed between the insulation vessel and the aluminum containers to intercept the 
residual heat losses through the insulation walls thus avoiding the boiling of the LAr bulk. 
Circulation of 2-phase nitrogen, with its latent heat, was chosen instead of under-cooled single phase liquid with much lower 
specific heat as used in Pavia test run. This new solution allowed using lower power circulation pumps and guaranteed fast 
cooling-down phase while guaranteeing thermal gradients within specification. Moreover it forced de-stratification of LAr 
during normal operation maintaining uniform and stable temperature in the LAr bulk. 
For safety reasons the cooling shield  was designed to operate also with liquid nitrogen circulation driven by gravity 
without the necessity of a dedicated circulation pump thus ensuring the cooling even in case of emergency 
situations such as the total lack of power supply.

\begin{figure}[htbp]
\centering
\includegraphics[width=15cm]{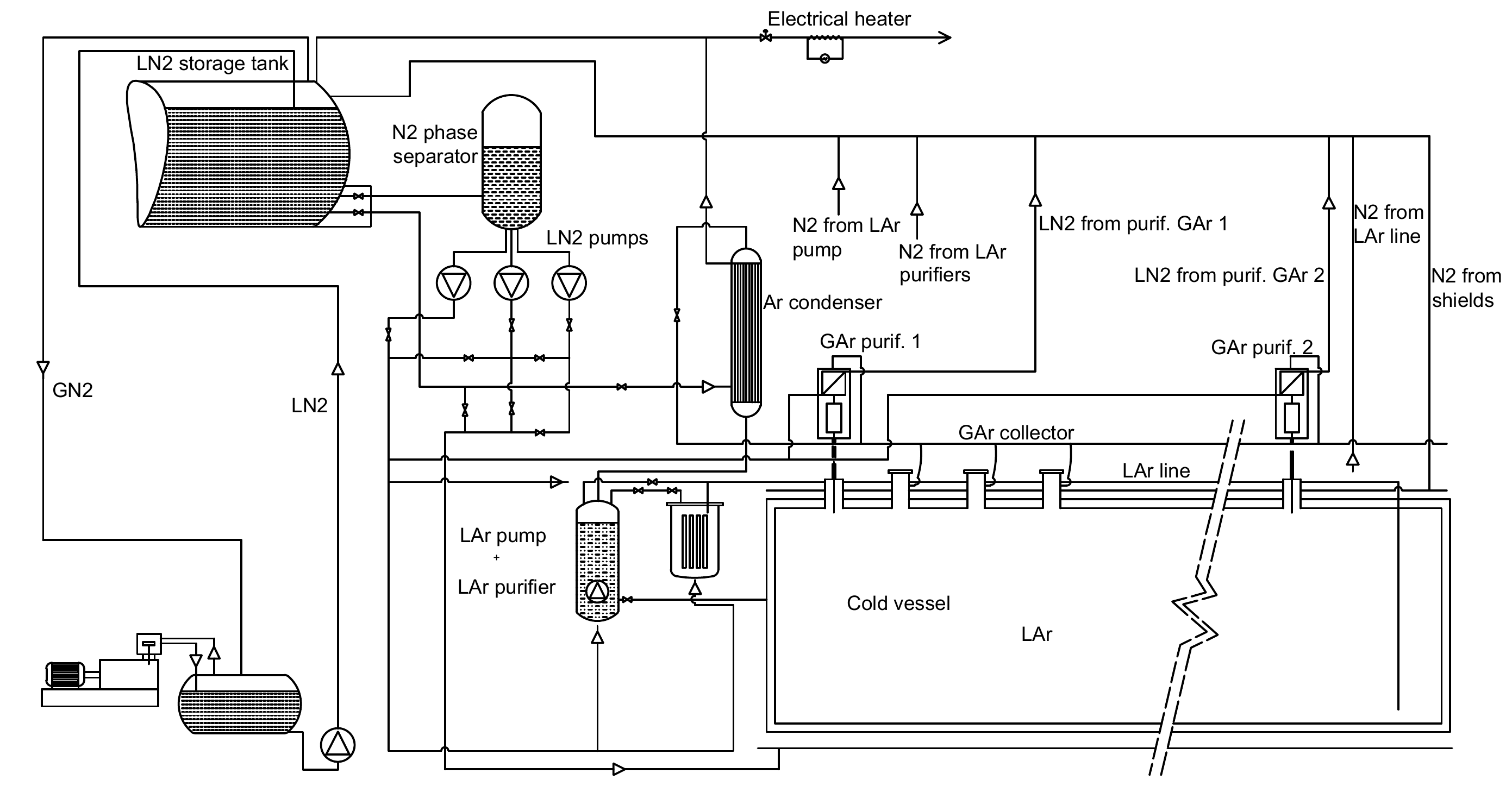}
\caption{Schematic view of the ICARUS T600 cryostat with argon and nitrogen circuits including the implementation 
of the system to operate in full gravity-driven mode even for GAr re-condensers.}
\label{cryots}
\end{figure}

Each T600 module was equipped with two gas and one liquid re-circulation systems to  purify gas top 
and liquid bulk respectively, in order to be used to reach and maintain the required LAr purity for the physics run. 
A schematic view of all the argon and nitrogen cryogenic circuits is shown in Fig.~\ref{cryots}.

The gas re-circulation system was intended to purify the internal gas phase since the initial filling phase when the 
outgassing rate was still present (it decreases exponentially with temperature) and to act as detector pressure stabilizer 
during steady state operation. The gas re-circulation units collected Ar gas (GAr) from the chimneys hosting the read-out 
cables and the feed-through 
flanges. GAr on the T600 top is warm and dirtier with respect to the liquid, as it is in contact with hygroscopic 
plastic cables and it could be polluted by possible small leaks due to the presence of several joints on each chimney.
The gas was re-condensed and then dropped into a liquid nitrogen cooled Oxysorb\texttrademark filter placed below 
the re-condenser. Finally the purified LAr flowed back into the LAr bulk just below the liquid/gas interface. The
 condenser was fed with liquid nitrogen at the temperature required for efficient re-condensation of the argon gas, 
 by means of forced circulation. The argon re-circulation rate was normally kept at the maximum rate of 25 GAr Nm$^3$/h/unit. 

The re-circulation in  liquid phase was instead devoted to massively purify LAr and  to reach and maintain 
the highest purity level after the cryostat filling and, in addition, to rapidly restore argon purity in case of  
accidental pollution during the detector operation. Each LAr re-circulation system extracted LAr at about 2 m 
below the surface on the 4 m height sides of the 
T600 modules and injected it on the opposite side, 20 m apart, close to module floor, through a horizontal pierced 
pipe that ensured a uniform distribution over the vessel width.
Each system was equipped with an immersed cryogenic pump (ACD CRYO AC-32 centrifugal pump) placed inside an independent 
dewar. From the pumps reservoir, the circulated LAr went  through a battery of four 
Oxysorb/Hydrosorb\texttrademark filter cartridges (connected in parallel) and was then re-injected into the detector volume. 
Each set of filters had a nominal O$_2$ absorption capacity exceeding 200 normal liters, sufficient to purify a module 
starting from standard commercial liquid argon (O$_2$ concentration $\approx$ 0.5 ppm). The maximum re-circulation rate 
was $\approx$ 2 m$^3$/h,  resulting from the pump throughput and the filter battery impedance and corresponding to a full 
volume re-circulation in about six days. Liquid nitrogen was used to cool  the pump vessel, purifier cartridges and all 
the Ar transfer lines.

Three Barber Nichols external motor centrifugal pumps BNCP-51B-000 model were installed to circulate liquid nitrogen 
inside the T600 cryostat circuits: one was dedicated to the cooling shield and one to the argon re-circulation systems, 
while the third one was redundant and ready to start in case of need. Each pump was located inside an independent 
cryostat fed by gravity from the nitrogen phase separator connected to the main liquid nitrogen storage on the top 
of the supporting structure. One more extra spare pump was present on site.

The two-phase nitrogen returning from cryostat screens and from the Ar gas and liquid Ar re-circulation system cooling 
was sent back to the two 30 m$^3$ liquid nitrogen storages\footnote{In steady state conditions both the 30 m$^3$ reservoirs 
 were dedicated to nitrogen storage (filled up to about 80 \%), while during commissioning one was used for liquid argon.} 
 on the top of the ICARUS service structure.
For safety reasons connected with the long term operation in underground the T600 nitrogen circuit was designed to work 
in closed loop by means of a dedicated nitrogen re-liquefaction system. 
Emergency operation in open circuit was also possible and easily handled in case of prolonged stops of the system 
provided that liquid nitrogen reservoir was maintained through periodic refill by trucks.

The nitrogen re-liquefaction system was dimensioned to cover the nominal total cold power required for the whole 
ICARUS T600 plant with at least 50\% margin, determined by the design heat load through the insulation (including 
the heat input through joints, cryostat feet and cables), the foreseen nitrogen consumption for the cooling screen
 (included the circulation pump and distribution lines) and the Ar gas and liquid Ar re-circulation-purification systems.

\begin{figure}[htbp]
\centering
\includegraphics[width=15cm]{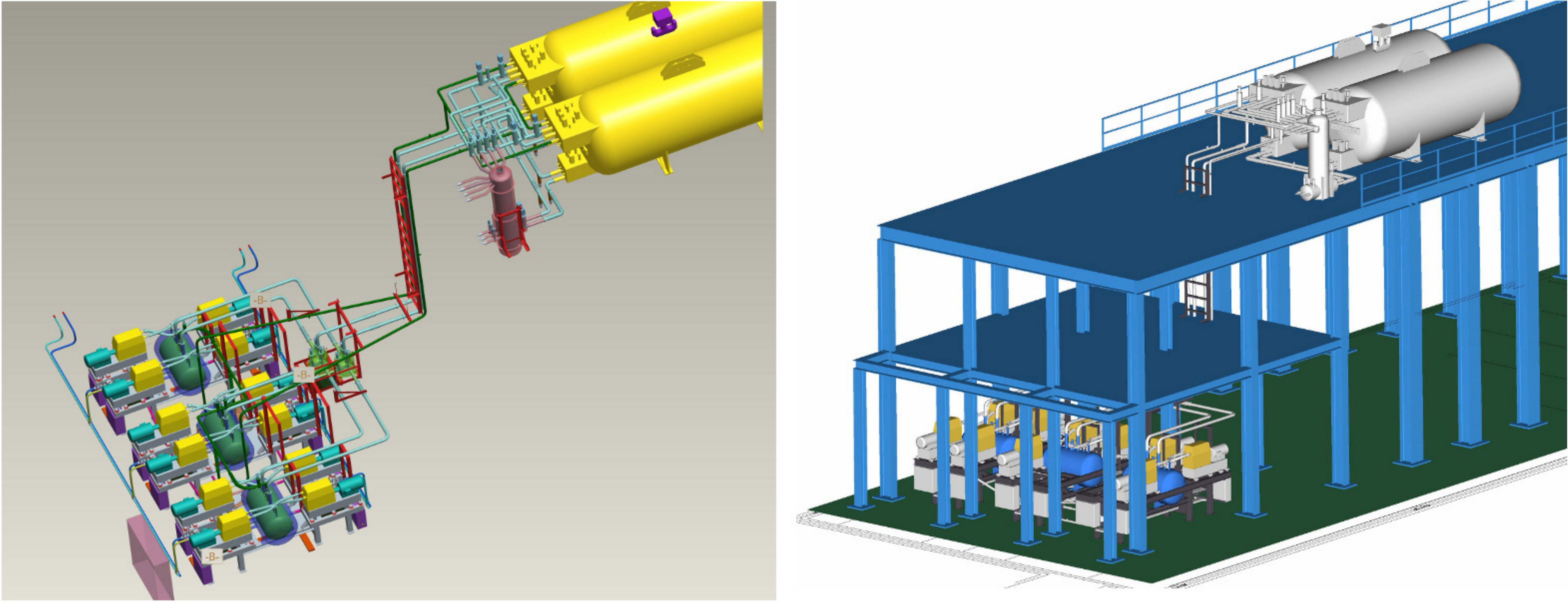}
\caption{Schematic view of the ICARUS T600 LN$_2$ re-liquefaction system plant composed by the 12 cryo-generators 
installed on the Hall B floor and the two storage tanks on the top of the ICARUS supporting structure. A dedicated 
2 t crane was installed above the skids to ease cryocooler maintenance.}
\label{stirling}
\end{figure}

The implemented system consisted of twelve Stirling\footnote{www.stirlingcrogenics.com} Cryogenics BV SPC-4  
(4-cylinder) cryo-coolers\footnote{At the time of the commissioning only ten cryo-coolers were installed. 
The system was later upgraded to guarantee larger redundancy.}, based on the ``reverse Stirling thermodynamic cycle'', 
delivering 4.1 kW of cold power each at 84 K with an efficiency of 10.4\% (each unit requires 45 kW power for its electrical motor).
This system was organized in 3 skids, each one composed by 4 cryo-coolers  and one 1 m$^3$  reservoir for liquid nitrogen (LN$_2$)
(Fig.~\ref{stirling}). 

During steady state the liquid nitrogen tank pressure was kept stable by the re-liquefaction system:  the typical 
working  pressure was  $\approx$ 2.1 bar abs  (corresponding to about 84.5 K temperature) in order to maintain 
liquid argon at about 87.5 K with an overpressure of about 100 mbar. Nitrogen gas present in the two 30 m$^3$ LN$_2$ storage tanks 
was re-condensed in the 1 m$^3$  reservoirs 
and then  injected in one of the two 30 m$^3$ storage tanks by means of two redundant cryogenic transfer pumps 
(Barber Nichols external motor centrifugal pump BNCP-68-M1 model).  

All units operated independently, automatically switching on/off  to keep the nitrogen pressure at any given 
set-point thus delivering the actual cold power needed by the system. This design provided  large flexibility 
in delivering cold power because of its intrinsic factorization; in addition it featured further advantages such 
as less critical maintenance stops (one unit at the time) without interference with the continuous cooling demand, 
electrical consumption minimization, easy plant expansion.

\subsection{ICARUS T600 safety equipment, control systems and infrastructures}
\label{T600safety}

The ICARUS T600 detector was equipped with several intrinsic safety systems mainly dedicated to prevent and eventually 
confine possible liquid and gas spillages:

\begin{itemize}
\item[-] each T600 module was protected by two ADAREG$^{TM}$ magnetic disks opening at an internal relative pressure of
 0.45 mbar and closing when the relative pressure drops below 0.4 mbar; three manual valves were also installed 
 on three different feed-throughs chimneys. They could be operated manually to lower the pressure;
\item[-] the volume between the insulation vessel and the two T600 modules was monitored by temperature sensors 
and protected by two safety magnetic disks and one safety valve;
\item[-] all the magnetic disks outlets were connected to a  system of two batteries of  three passive heaters 
(filled by half-rings of Steatite with high thermal exchanging surface) that bring the exhausted argon to room 
temperature (Fig.~\ref{t6003d}). A blower and  a 10 kW electrical heater were installed to warm-up the saturated 
battery while the other one was in operation;
\item[-] the aluminum honeycomb walls of the T600 modules were monitored and protected against pressure increase;
\item[-] redundant safety valves for all the cryogenic tanks and pipes were installed;
\item[-] all the possible points of exhaust (safety valves and rupture disks) were collected together into a single 
vent line to convoy gas to the ventilation extraction port; this vent line was provided with an electrical/pneumatic
 control to vent at a fixed set point;
\item[-] a 50 kW electrical heater (regulated by the temperature at the exit)  to warm-up all the cold gas exhaust 
was put after the discharge valve before the extraction port of the ventilation system in the Northern side of the
 Hall B, see Fig.~\ref{t6003d}. The discharge valve plus the electrical heater were used to vent warm nitrogen from 
 the  two 30 m$^3$ tanks in case of emergency when the nitrogen  re-liquefaction system was off (open-loop operation)
  or to vent the transfer line during the liquid refill of the storage vessels.
\end{itemize}

The whole T600 plant (argon purification and nitrogen circulation systems) and the nitrogen cryo-coolers system, 
were provided with two local and independent control systems. Both systems were based on widely employed industrial 
devices (Allen-Bradley for the T600 cryogenic plant and Hitachi for the cryo-coolers). A high level of redundancy
was achieved with automatic intervention to guarantee the maximum operational continuity. 
All the plant parameters of the T600 cryostat were handled by two redundant PLCs (Programmable Logic Controllers) and  
the critical nodes related to safety were also connected  to a third PLC. Automatic process control was developed in 
order to promptly react to any parameter change or emergency.  A common interface  based on a SCADA server 
(iFix Intellution installed on an industrial PC) was also available for higher level, remote, supervision and control system. 
to record and store all the relevant parameters and events and issue alarms and automatic notifications. 
It was interfaced with the general LNGS underground Safety Control Room.

The ICARUS Area was equipped with several safety sensors and devices, also monitored with the same SCADA supervision system:

\begin{itemize}
\item[-] 13 temperature sensors (PT100 heads with range from -50$^o$C to +150$^o$C) located near the floor all around 
the T600 plant to detect temperature decrease connected to accidental cold gas or cryogenic liquid leaks;
\item[-] 13 oxygen sensors (Draeger-Politron II-O$_2$) located  near the floor all around the T600 cryostat and the 
nitrogen re-liquefaction plant, 3 on the other two supporting structure levels and other 4 near the ICARUS Control 
Room and the cryogenic liquid downloading station, to promptly detect an oxygen concentration decrease due to accidental 
cold gas or liquid leaks;
\item[-]  an on-line smoke monitoring system and aspiration ports for each electronic rack;
\item[-] an on-line smoke monitoring system in the ICARUS Area for fast fire detection;
\item[-] a closed circuit TV system composed by 9 cameras located all around the T600 plant.
\end{itemize}

All the possible emergency situations were extensively studied and several scenarios were identified and classified in 
terms of  risk level. Significant spillage of cold liquid and/or gas was defined as the most critical scenario (cryogenic emergency).
Dedicated infrastructures with high redundancy were implemented for the whole apparatus:

\begin{itemize}

\item[-] Hall B air extraction system, to  create appropriate differential pressure between the Hall B and the rest of 
the underground Laboratory.
During normal operation,  about 7,000 m$^3$/h air were inlet in the Southern end of the Hall B and extracted (from bottom 
and top) in the North end side, maintaing a slight overpressure with respect to the rest of the Laboratory.
In case of ``cryogenic emergency'' the extraction system was set to create  in the Hall B a lower pressure with respect
 to the rest of the Laboratory;

\item[-] an emergency extraction system made of capillary pipes  extracting gas from the bottom level of  ICARUS area 
and connected to the main Hall B air extraction system. It  protected the personnel and the plant in case of cryogenic 
liquid and cold gas spillages; in this configuration the other extraction ports of the Hall B are closed and the only 
extraction way is through this emergency system;

\item[-] a redundant electrical plant (including a spare source and double distribution line) consisting of a 850 kW 
electrical cabinet dedicated to the T600 plant and of a second distribution system powered by an independent LNGS 
cabinet and distributing electrical power to the  T600 using an alternative geometrical path in order to enhance reliability;

\item[-] an uninterrupted power supply (UPS) for control and safety systems, detector relevant components and for the 
ICARUS control room;

\item[-] a water cooling system to cool the re-liquefaction plant equipped with redundant pumps to enhance water
 pressure (5 bar). The typical required water quantity per unit was 60 l/min;

\item[-] a redundant compressed air system to actuate pneumatic or electro-pneumatic valves.  To enhance redundancy 
an extra compressor ready to start and several nitrogen gas  bottles were installed;

\item[-] a manually activated diesel generator covering the base electrical power needs of the cryogenic plant 
(for valves, sensors, controls, electrical heaters for argon and nitrogen exhaust, nitrogen pumps, insulation vacuum pumps).

\end{itemize}

\subsection{The slow control system}
\label{slowcontrol}

The behavior of the two T600 modules  was continuously monitored during the whole critical phases of evacuation, 
cooling and filling with liquid Ar,  through the survey of dedicated sensors, installed in each module for this purpose:

\begin{itemize}
\item[-] 8 sensors to measure the module inner wall displacement during the vacuum phase and 10 potentiometric 
linear sensors to monitor the mechanical behavior of the insulation walls under differential pressure and thermal gradients;
\item[-] 4 sensors to control the rotation speed and the adsorbed current of the 4 turbo-molecular pumps 
used during the vacuum phase;
\item[-] 2 internal pressure sensors and one external pressure sensor;
\item[-] 30 platinum resistors (Pt1000 type in West T600 module and Pt10000 type in the East one) for internal 
temperature measurement and 40 external platinum resistors Pt1000 located on the outer insulation surface to monitor 
thermal losses;
\item[-] 14 capacitive position meters to measure the movement of the springs,  that compensate the thermal 
contraction of the wires during the cooling phase;
\item[-] 16 continuous  level sensors to monitor the liquid Ar filling and 20 carbon resistors, acting as
 level probes, to monitor the final part of the liquid Ar filling and precisely set the final level.
\end{itemize}

\noindent

The acquisition and storage of the slow control signals  was based on two National Instruments\footnote{www.ni.com} 
compact Field Point modules, one for each T600 module.

\section{Commissioning}
\label{commissioning}

The commissioning of the ICARUS T600 cryogenic plant started at the beginning of 2010 following the same approach 
adopted in the successful surface test run in 2001~\cite{t600}, with an extra attention to maximize safety and minimize 
interference with other underground activities.
The procedure consisted in four main subsequent phases: (i) detector volume evacuation, (ii) cryostat cool-down, (iii) 
liquid Ar filling and GAr purification/recirculation start-up, (iv) LAr-TPC detector commissioning and LAr 
purification/recirculation start-up. The most critical phases were  remotely operated and controlled from the ICARUS Control Room located in the Southern side of Hall B. 

A dedicated area to unload cryogenic liquids  from trucks was set-up in the Hall B in correspondence of the
 LNGS Truck Tunnel. Vacuum jacked liquid nitrogen and liquid argon transfer lines were installed from the unloading 
 station to the two storage tanks on top of the ICARUS service structure at a distance of about 80 m.

\subsection{Vacuum phase}

The adopted strategy to ensure an acceptable initial LAr purity relied on the cryostat evacuation down 
to a residual pressure of about 10$^{-4}$ $\div$ 10$^{-5}$ mbar to perform an appropriate out-gassing of 
all the internal walls and detector materials and to remove air pockets in the inner detector structures.

To this purpose, each T600 module was equipped with four identical remotely controlled pumping groups, 
mounted on four UHV-CF200 flanges on the T600 insulation top. Each pumping system consisted of 
a 24 m$^3$/h primary Varian\footnote{www.varianinc.com} Dry Scroll DS600 pump, a 1000 l/s Varian Turbo-V 1001
 Navigator pump and three electro-pneumatic gate valves (two on UHV-CF200 flanges, one on UHV-CF35 flange), 
 which allowed to intercept the pumping group, isolate the inner volumes and start vacuum phase with only primary pumps. 
 A  safety valve was mounted in parallel to each dry scroll pump to prevent air return in case of power failure. 

Before evacuation, the tightness of  both T600 modules was tested to a moderate internal over and under- relative 
pressure\footnote{Ambient pressure in the tunnel is about 900 mbar as it is located at the height of about 1000 m above the see level.} 
in order to find and eventually repair major leaks.
The commissioning procedure continued with the vacuum pumping of all the volumes to be filled with argon both in liquid and gaseous phase 
(the main volumes, the purifiers, the argon transfer lines and the recirculation units) in order to remove air and other pollutants. 
In order to limit vacuum load only to the external skin of the cold vessels  panels, the aluminum honeycomb structures of the cold vessels 
walls were evacuated by means of rotary vane pumps and their pressure was continuously monitored.

\begin{figure}[htbp]
\centering
\includegraphics[width=14cm]{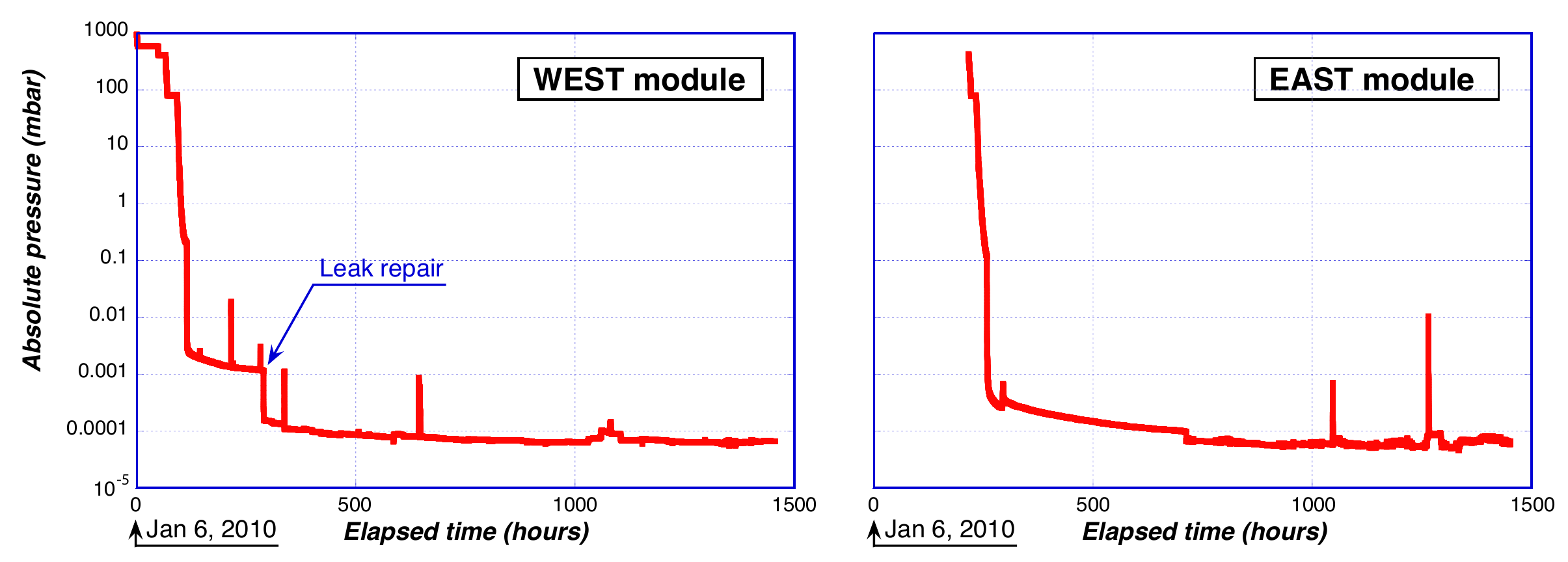}
\caption{Pressure on the West (left) and East (right) modules as a function of the pumping time. The spikes in the plot are due to
interventions devoted to leak repair. The large step at about 300 h in the West module is due to the repair of a major leak. Peaks are due to the stop of one of the four turbo-molecular pumps.}
\label{vacuum}
\end{figure}

The evacuation of the two modules  was performed in sequence. In both cases the effective time to reach 0.2 mbar was approximately 30 hours.
 The turbo-molecular pumps were then switched on to proceed with the high vacuum phase. 
For both modules a systematic search and repair of leaks resulted in a sudden improvement of vacuum 
level. Fig.~\ref{vacuum} shows the pressure evolution in the two modules during the pumping phase as a function of the effective pumping time. 

During the whole vacuum phase a continuous monitoring of the mechanical deformations of the 
inner walls  was carried out by the eight position meters (see Sec.~\ref{slowcontrol}). 
 As expected from simulations and from the Pavia experience, the walls deformation increased 
 linearly with decreasing pressure and reached a maximum of about 35 mm at the center of the 
 longest vertical walls in both modules.

The target pressure of 10$^{-4}$ mbar was reached in less than three weeks in both modules; 
then vacuum pumping  was continued for a period of three months before starting the cooling phase. 
The final equilibrium pressure  was  4.5 $\cdot$ 10$^{-5}$ mbar (3.8 $\cdot$ 10$^{-5}$ mbar) for the 
West (East) module. These values correspond to a global leak rate of about
 6 $\cdot$ 10$^{-2}$ mbar l/s (4 $\cdot$ 10$^{-2}$ mbar l/s), dominated by internal outgassing 
 as verified by measuring the residual gas composition with mass spectrometers 
 installed on two of the top flanges. The measurements showed a $1\div 10$ relative content of 
 air to water, showing that outgassing was the dominant source of the residual gas which was 
 expected to freeze on the internal surfaces during the cooling-down phase; contributions from 
 other components were negligible.

\subsection{Cooling phase}

The Stirling plant was commissioned in advance and it had been successfully operating for one year when the cooling phase started. 
During this period, a series of tests was performed, simulating different working situations, transient phases and cold power
requests up to a maximum of 36 kW and demonstrating the correct behavior of the system in agreement  with specifications. 
Failure tests were also successfully performed, including lack of water cooling,  power-cuts, liquid nitrogen transfer pump stop. 

Immediately after vacuum pumping was stopped, the two T600 modules were loaded with ultra-pure gas argon 
(Ar N60: $<$ 0.5 ppm H$_2$O, $<$ 0.1 ppm O$_2$, $<$ 0.3 ppm N$_2$) at 100 mbar overpressure, to minimize back-diffusion of air from residual leaks.
Then LN$_2$ circulation started inside the cooling screens using nitrogen from the 30 m$^3$ storage tank. Both forced  and gravity driven circulations 
were successfully tested and operated. A constant overpressure of 100 mbar was maintained by means of continuous injection of purified gas argon in both modules.
Cryostat internal pressure, cryostat wall displacement, temperature gradients on the wire chambers, insulation external temperatures and displacement were  monitored.

\begin{figure}[htbp!]
\centering
\includegraphics[width=8cm]{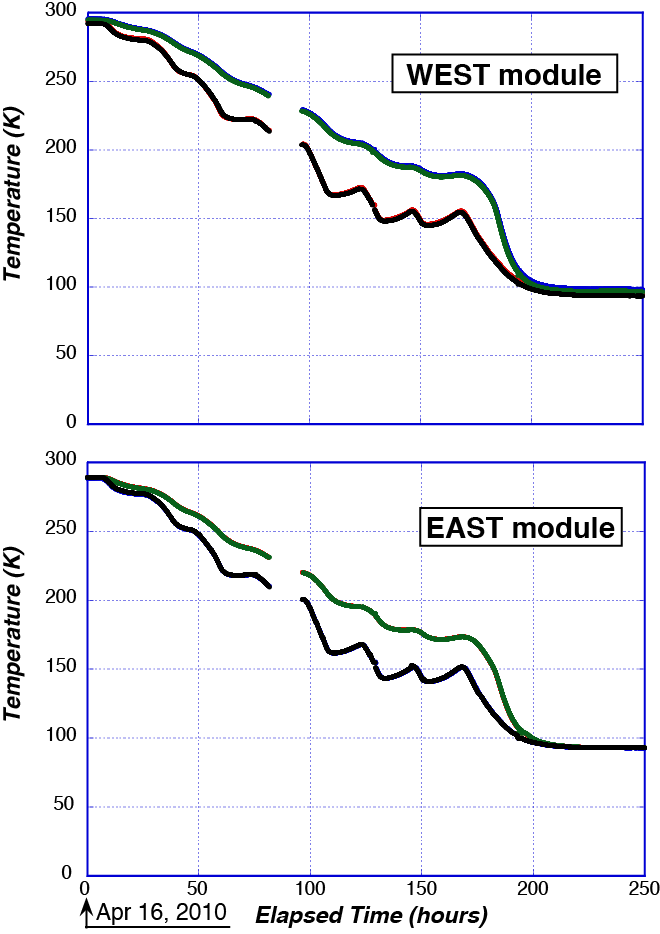}
\caption{Internal temperature trend on the wire chambers structures in the West (top) and East (bottom) modules as a function of time along the whole cooling phase. 
The values of two temperature probes (out of 15) for each wires chamber are shown, one on the top and one on the bottom of the structure.}
\label{tempwest}
\end{figure}

During the cooling phase, all cryo-coolers were fully active and able to handle most of the nitrogen evaporation, which was partially 
exceeding the re-condensation power only during the beginning of the cooling of the ~100 ton mass of the metallic containers.  
The residual nitrogen vapor was warmed-up through the 50 kW electrical heater and safely evacuated from Hall B via the ventilation system.  
The cooling phase lasted about eight days, reaching 90 K at an average rate of about -1 K/h. The cooling was slowed-down when required to 
keep the internal temperature gradients on the wire chambers within specifications (50 K). Fig.~\ref{tempwest} shows the temperature 
trend on the wire chambers structures of the West and East modules; the wiggles on the cooling trends are due to the stops of the liquid nitrogen
circulation that were used to keep the thermal gradients within specifications. The total 
LN$_2$ consumption was only 55,800~l, 
to be compared with an estimate of 200,000~l required in case of full open loop.

\begin{figure}[htbp]
\centering
\includegraphics[width=11cm]{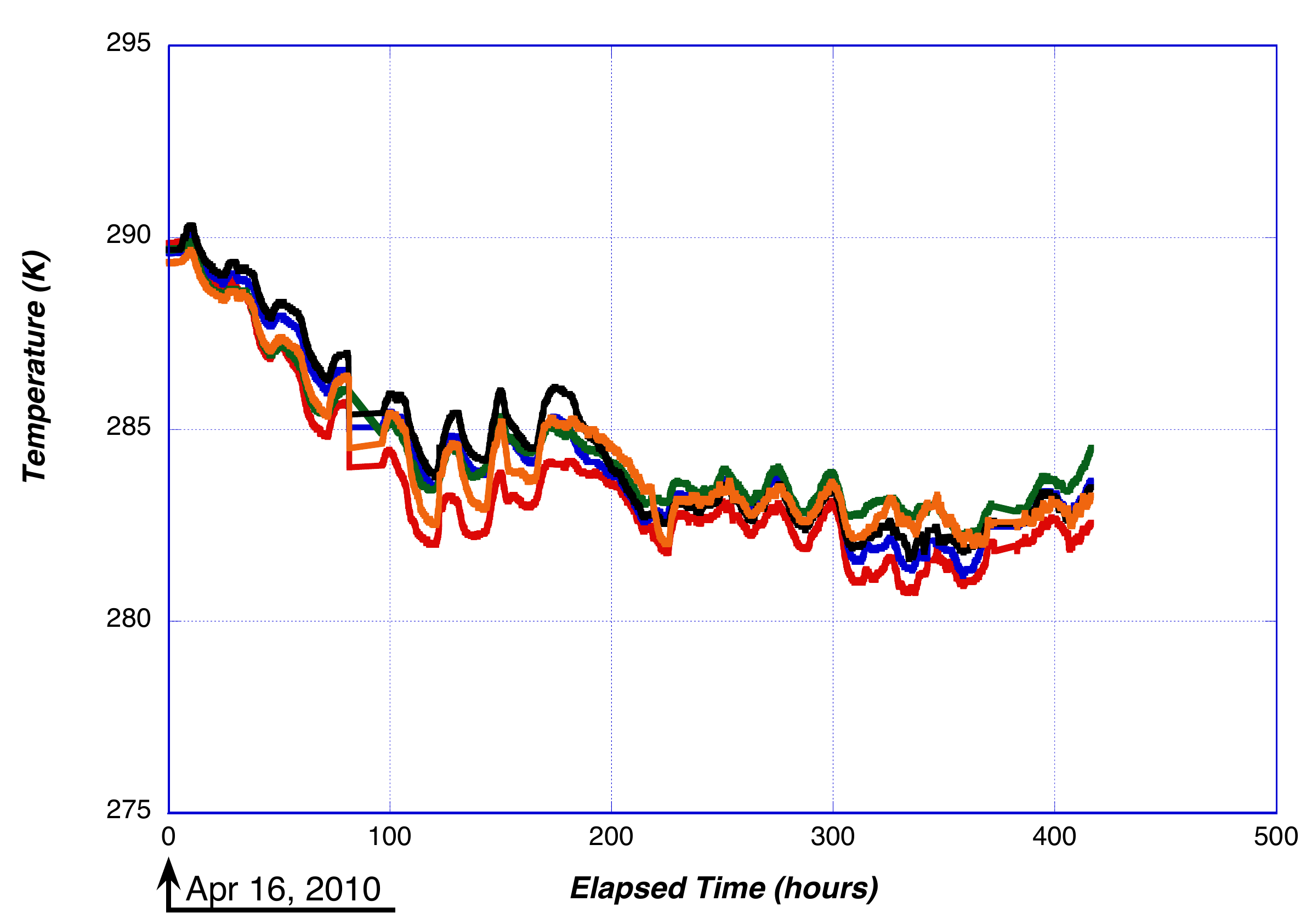}
\caption{Temperature trend of Pt1000 probes located on the outer skin of the insulation vessel: South (red), 
West (blue and green), East (black and pink). Temperature values stabilization were 282 K $\div$ 284 K while the difference 
with ambient value was  about -7 K.}
\label{outresume}
\end{figure}

After the conclusion of the cooling phase, when the stabilization of the system was achieved and with nitrogen re-condensation system off, 
the power consumption (insulation losses, plus feet, pipes, cables, chimneys heat input) was determined to be 24 kW in total (corresponding 
to about six active Stirling units), well within the capability of the re-liquefaction system with all the cryogenic plant activated. 
The specific contribution to the heat losses of the thermal insulation was estimated to be $\sim$ 20 kW as derived from the temperature 
differences with respect to ambient air measured on the outer skin of the insulation panels (see Fig.~\ref{outresume}).

The insulation bottom panel resulted  within specifications with an internal working pressure of the order of 10$^{-4}$ mbar.

\subsection{LAr filling phase} 

To ensure filling without interruptions, one of the 30 m$^3$ tanks was used as LAr storage buffer, 
 continuously feeding the purification cartridges of re-circulation systems of the two modules.

A dedicated cryo-cooler  (Stirling Cryogenics BV 1-cylinder SPC-1 500 with a nominal cold power 
of 1 kW at 77 K) was installed on the liquid argon tank only for the commissioning phase with the 
aim of stabilizing the argon pressure by means of re-condensation.

Each liquid argon delivery (13 m$^3$), certified to be within purity specifications (H$_2$O $\leq$ 1 ppm, O$_2$ $\leq$ 0.5 ppm, N$_2$ $\leq$ 3 ppm),
was downloaded into the LAr 30 m$^3$ vessel. An additional buffer tank (1.3 m$^3$) was installed at the liquid argon unloading station, to keep the 80 m 
long LAr transfer line cold and to preserve argon quality between two consecutive deliveries.

A filter composed by a standard Oxysorb$^{TM}$/Hydrosorb$^{TM}$ cartridge with high purification capability  was put at the outlet of the 30 m$^3$ vessel, 
to avoid early saturation of the T600 main purification cartridges. The liquid argon quality was monitored on-line at the 30 m$^3$ storage inlet and 
downstream this filter by means of a gas chromatograph. On average, a typical contamination of 30 ppb in oxygen and 100 ppb in 
nitrogen\footnote{Initial nitrogen content in LAr was not removed by means of the ICARUS purification system. Even if nitrogen 
is not electro-negative, it has to be maintained as low as possible as it affects the scintillation light production, that is 
fundamental for internal trigger and timing purposes as explained in the following.} after the additional filter was measured. 

To minimize argon pollution due to outgassing from surfaces, T600 filling was performed in the shortest possible time, 
while avoiding to reach the opening pressure of the magnetic safety disks. Before starting the filling phase, 
all transfer lines and storage vessels were  accurately purged with high quality argon, until the residual oxygen and water 
content was  lower than 1 ppm.

In order to intercept  residual outgassing impurities, the four Gar recirculation/purification units were put 
into operation at the beginning of the filling, at a rate slightly exceeding the nominal value of 25 Nm${^3}$/hour per unit.
To guarantee an internal overpressure, the cryostat filling was started with ultra-pure Ar gas followed by injection of the 
first 10,000 liters of liquid argon in West module. A similar procedure was followed for the East cryostat.

\begin{figure}[htbp]
\centering
\includegraphics[width=14cm]{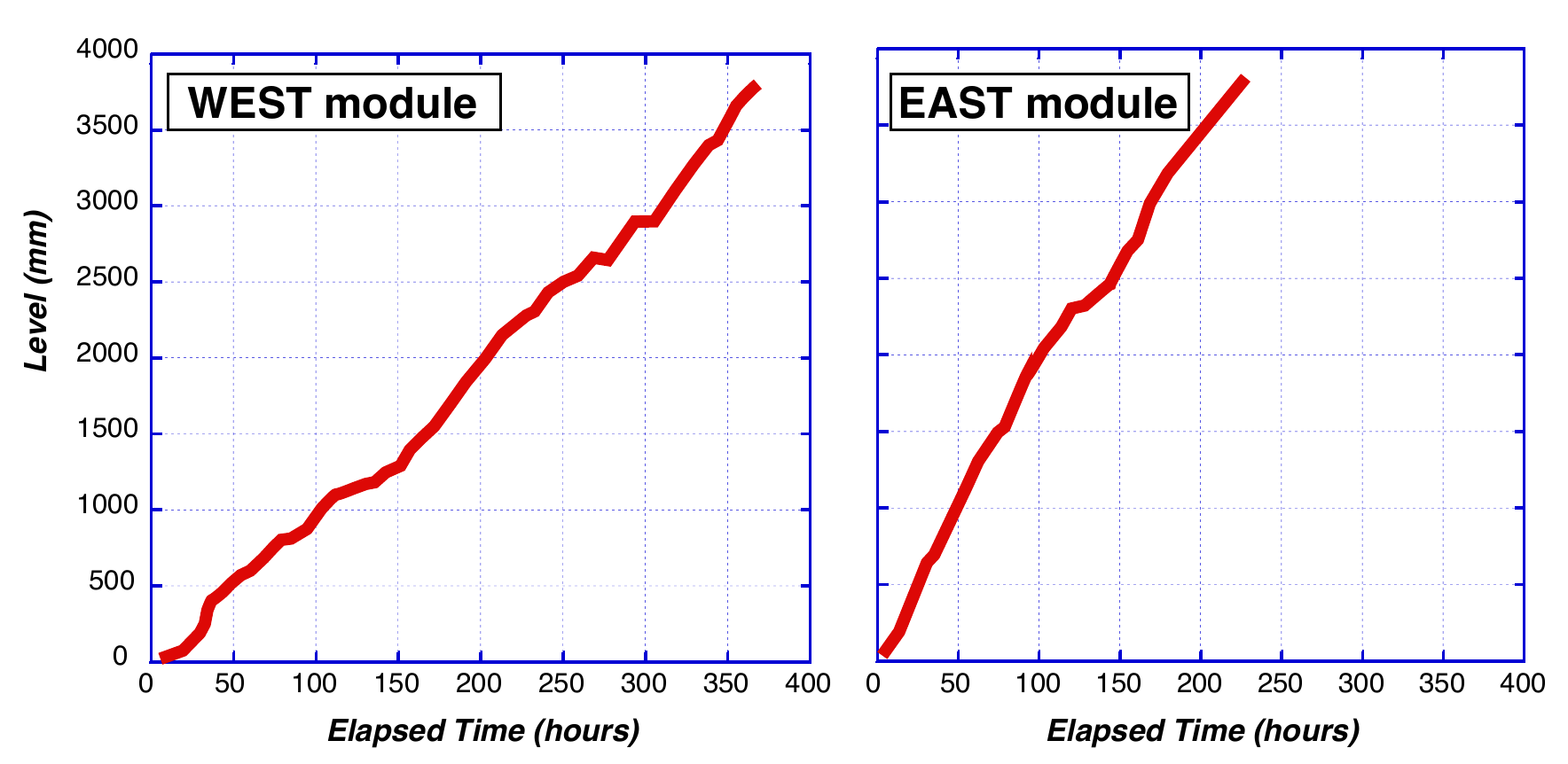}
\caption{Liquid argon level during filling inside West (left) and East (right) modules. }
\label{filling}
\end{figure}

Few days after, as required to fully thermalize the inner detector structures, the continuous liquid argon filling in both 
modules started with an overall rate of about 2 m$^3$/h.  The whole filling lasted about two weeks and was carried 
out without the need of opening the cryostat exhaust valves.
The final level of liquid argon was precisely set at 3825 $\pm$ 5 mm from the cold vessels floor by means of the arrays 
of discrete level meters described in Sec.~\ref{slowcontrol}, to ensure a safe coverage of the high voltage region of
 the TPC field cage. The total amount of downloaded liquid argon was 610,511 liters (47 deliveries).
In Fig.~\ref{filling} the liquid argon level trend during filling is shown for both modules, as monitored by
 the continuous level meters described in Sec.~\ref{slowcontrol}.

\subsection{Detector commissioning and LAr re-circulation and purification start-up} 

After the completion of the cryogenic plant commissioning, the T600 detector steady state 
working conditions were reached in few days. Soon after, the TPC in the West module was 
activated by turning on the high voltage biasing system (- 75 kV at the cathode), 
the data acquisition and PMT trigger system: the first ionization track was immediately 
recorded and visualized~\cite{t600_jinst}. 

As already mentioned, the detector cooling and filling procedures did not produce any 
significant effect on the internal detector structures, the TPC wires and  the PMTs.
The initial electronic noise level was in agreement with expectations, without detectable 
microphonic effect due to the cryogenic plant operation.

\begin{figure}[htbp]
\centering
\includegraphics[width=14cm]{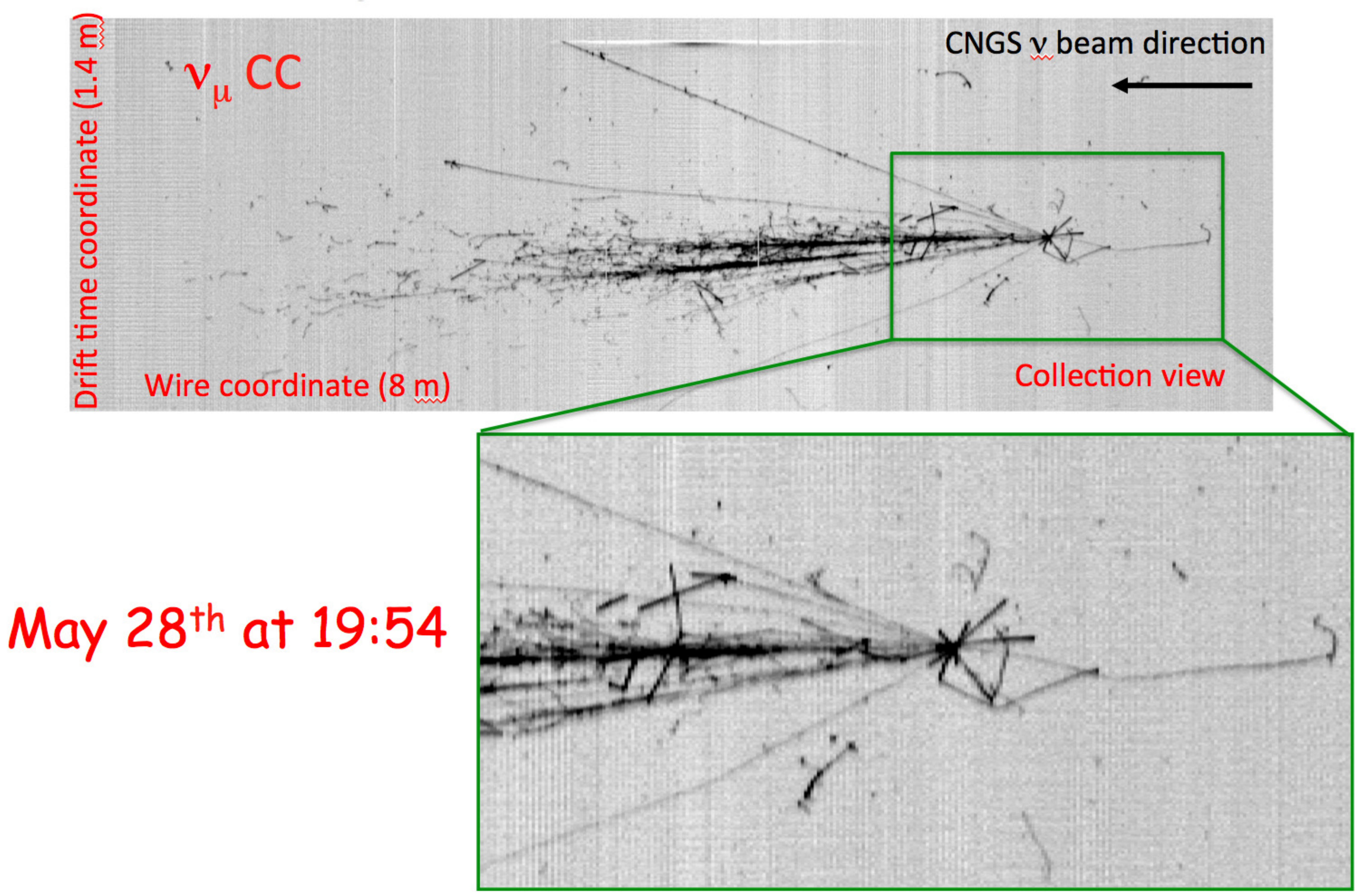}
\caption{First CNGS neutrino interaction observed in the ICARUS T600 detector.}
\label{firstevent}
\end{figure}

On May 28$^{th}$ 2010 the first CNGS neutrino interaction was recorded (Fig.~\ref{firstevent}). 
With uniquely the gaseous re-circulation systems active, the initial free electron lifetime  $\tau_{ele}$ was surprisingly measured with 
cosmic muon tracks to exceed  600 $\mu$s, uniform in the whole sensitive volume, corresponding to a liquid argon residual 
contamination of about 0.5 ppb of O$_2$ equivalent. 
Soon after, the East module was activated showing very similar initial performance. The liquid recirculation systems of both modules 
were also turned on, leading  to the steady increase of the LAr purity. 

\section{Cryogenic plant operation and performance}
\label{operations}

In the first few months of T600 operation, the whole cryogenic plant was tested, all the regulations fine-tuned  
and the related parameters were then stabilized.
A complete set of tests on possible failures of the apparatus and emergency events were satisfactorily carried 
out with the automatic intervention of the dedicated backup systems:
\begin{itemize}
\item[-] stop of LN$_2$ circulation pumps, replaced by redundancy or gravity driven operation;
\item[-] complete lack of electrical power, replaced by UPS and/or emergency diesel generator;
\item[-] lack of compressed air, replaced by operation with nitrogen gas bottles supply;
\item[-] failure of the main ICARUS PLC control system, recovered with automatic activation of redundancies;
\item[-] cold gas exhaust at the vent to check electrical heater functionality, temperature and oxygen trend at the exit.
\end{itemize}

Further redundancy and safety systems were afterward implemented  including a gravity driven 
LN$_2$/GAr heat exchange system to re-condense the GAr phase, an emergency power line, 
a fully pneumatic control system to operate valves, to specifically handle the emergency situations
 of total lack of power in the underground Laboratory. 

With all the implemented upgrades and the high intrinsic redundancy, the cryogenic plant was operated safely 
and reliably during the whole period of steady state T600 run even in case of severe emergency situations without 
stopping the data taking.  The control and supervision systems demonstrated to be extremely  efficient allowing 
adopting a smooth surveillance strategy based on an on-call group of experts intervening underground in case of need. 
As a result, the few recorded emergency situations, all to be ascribed to external power cuts, were rapidly and successfully handled. 

Major attention was dedicated to study the reliability and verify proper redundancy of the involved dynamic 
components, such as cryo-coolers and pumps, as typically represent the most critical part of a working plant. 

\begin{figure}[htbp]
\begin{center}
\includegraphics[width=14cm]{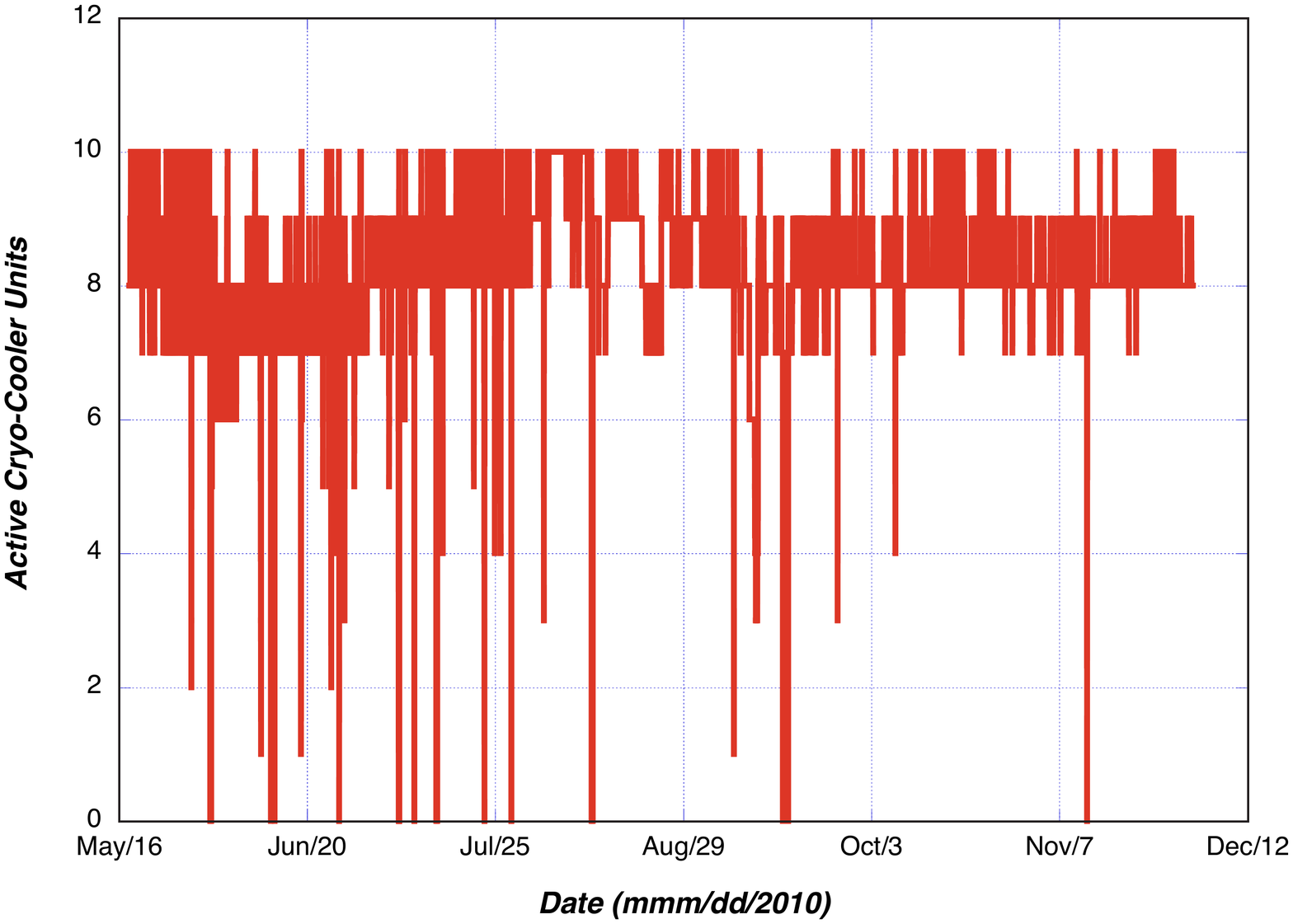}
\caption{Number of active cryo-coolers during the first months of the T600 operation. Some stops of the cooling system are present, due to power cuts or to failure tests.}
\label{NCryo}
\end{center}
\end{figure}

The  nitrogen re-liquefaction system demonstrated to efficiently cover the maximum T600 cold power 
request with flexibility and margin. The system  design  redundancy allowed to perform periodic 
maintenance to substitute some worn components (in the  3,000 $\div$ 12,000 hour range, depending 
on the device) without affecting the overall refrigeration capacity.
After few months of normal operation the average number of active cryo-coolers 
was found to be $\sim 9$ and never more than 10 over 12 units (Fig.~\ref{NCryo}).
Periodical measurements on all the cryocooler units allowed to determine the   
average effective cold power request of about 24 kW in agreement with the initial extimation after the plant commissioning.
The observed stability of the cold power request demonstrated that no aging effects or degradations of 
the thermal insulation performance occurred over the whole three years operation. The liquid nitrogen pumps 
(Barber Nichols Inc. with external motor) showed an extremely high reliability  
and stability without  interruptions over more than 10000 working hours.

The LAr recirculation system operation was found to be very effective in increasing the LAr purity which was continuously monitored by 
measuring the electron lifetime in LAr by the charge attenuation along ionizing cosmic muon tracks crossing the full drift volume. 
About 100 muon tracks were sufficient to measure day-by-day the electron charge attenuation within few percent.
The evolution of the residual impurity concentration was described with a simple model including the time 
needed to recirculate a full detector volume, a constant pollution rate due to external leaks/outgassing in 
gas phase and the initial contribution of internal residual outgassing which was assumed to vanish with 
time~\cite{t600_jinst}. Uniform distribution of the impurities throughout the detector volume was also 
considered, as experimentally supported by the lifetime measurement with muon tracks in different regions 
of the TPCs~\cite{T600_purity}. As result a full volume recirculation time of $\sim$6 days, in agreement 
with the nominal pump speed, and extremely low leak rates ($<$ few ppt/day O$_2$ equivalent) in both modules were found.
 
The ACD AC-32 immersed pumps resulted to be less reliable than expected due to excessive bearing case 
damages that caused frequent system faults with consequent purity drop. A precise and fast intervention procedure 
was adopted for the LAr pump substitution with a spare one. On average the time interval between two consecutive faults was about 2,000 h.

\begin{figure}[htbp]
\begin{center}
\includegraphics[width=15cm]{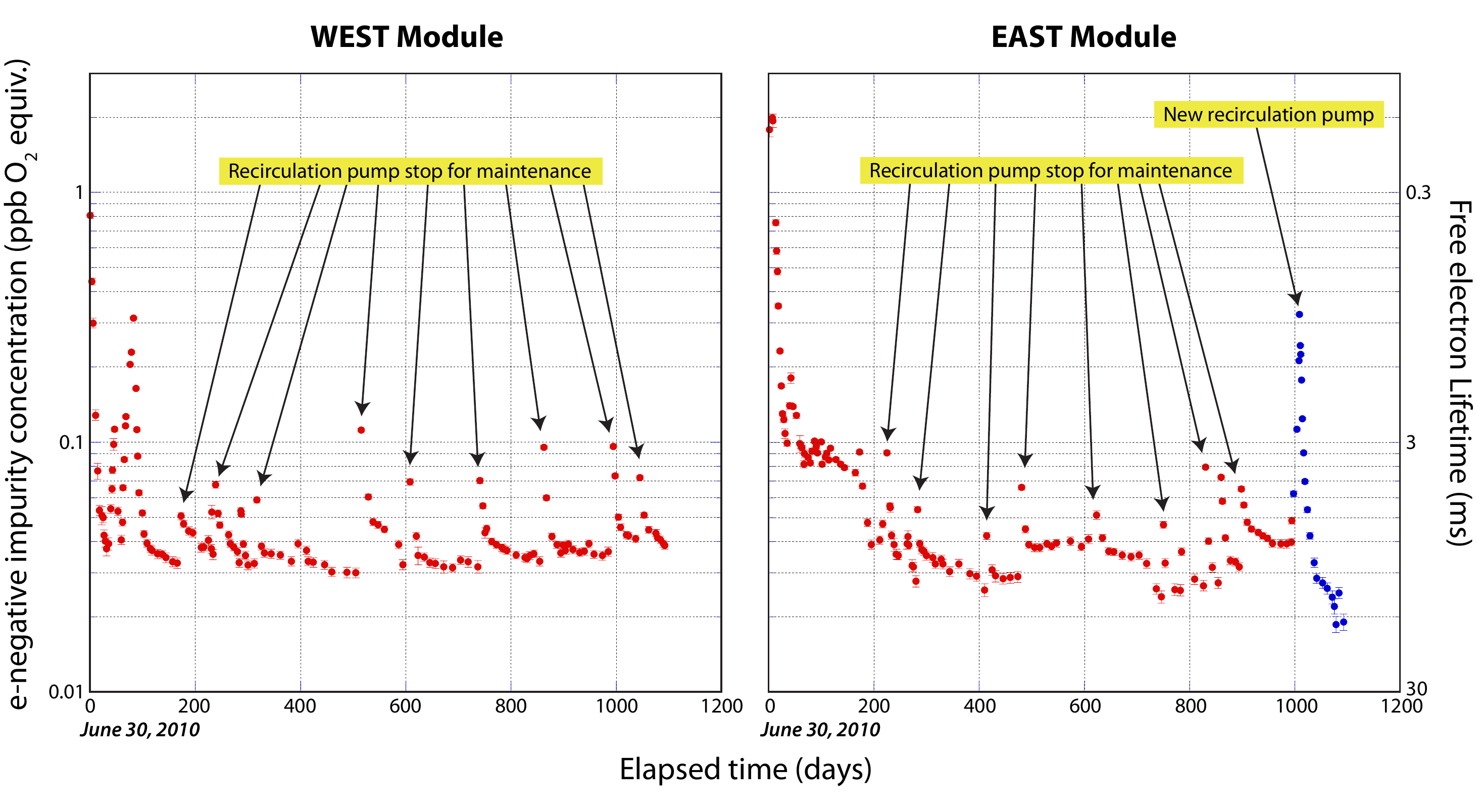}
\caption{Evolution of the concentration of electronegative impurities concentration in the West (left) and East (right) modules as a function of the elapsed time for more than two years of operation of the T600 detector. The corresponding free electron lifetime is shown on the right axis.}
\label{purity}
\end{center}
\end{figure}

In spite of the LAr recirculation stops, the impurity concentration was maintained below 0.1 ppb all over 
the detector run (see Fig.~\ref{purity}). Similar purity trend were observed in both the T600 modules. 
Electron lifetime values of the order of 7-8 ms were reached in both modules, corresponding to an impurity 
content of few tens of ppt that imply a maximum attenuation of free electrons of $\sim 12\%$.

In order to improve the reliability of the LAr re-circulation system a detailed study was performed 
comparing the different characteristics and reliability between the ACD 
immersed pumps used in the LAr circuits and Barber Nichols external motor ones present in the LN$_2$ transfer 
lines leading to a significant upgrade of the liquid argon recirculation system achieved in the last few months of 
detector operation~\cite{chiara_1}.
One of the ACD AC-32 pumps was substituted with a new Barber Nichols BNHEP-23-000 model similar to 
the other Barber Nichols pumps used on the liquid nitrogen circulation that showed much longer lifetime
between ordinary maintenance cycles. The new pump, characterized by 
magnetic coupling and vacuum housing,  was installed inside a new dedicated vacuum insulated cryostat.  
A LAr/LN$_2$ heat exchanger was added up-stream of the pump aspiration to under cool liquid argon and to 
ensure mono-phase liquid state. A Venturi flow-meter was inserted down-stream of the pump output, 
profiting of the mono-phase argon to measure the flux. After the new pump was switched on, the electron lifetime started 
increasing at a rate faster than before. 
At the end of the ICARUS data taking an electron lifetime exceeding 15 ms still rising was measured corresponding 
to  20 parts per trillion of O$_2$-equivalent contamination and an attenuation length of 25 meters, a milestone for 
any future project involving liquid argon TPC~\cite{T600_purity}. These results demonstrated the effectiveness of 
the single phase LAr-TPC detectors paving the way to the construction of huge detectors with longer drift distances. 
With the achieved purity level only $23 \%$ of the signal attenuation is expected at 5 m from the wire planes.

The oxygen contamination contents inferred by the electron lifetime measurements  all over the T600 detector run 
were perfectly compatible with the specifications for untouched scintillation 
light production and transport~\cite{O2contamination}\footnote{O$_2$ contamination in LAr leads to the attenuation 
of both the free electron charge (via attachment process) and the scintillation light (via quenching and absorption mechanisms).
The request on O$_2$ concentration to avoid scintillation light reduction is much less stringent than that for electron 
attachment (effects are visible for concentrations above 0.5 ppm).}. Together with oxygen, also the nitrogen concentration had to be 
continuously monitored as it was demonstrated that 
concentrations of few ppms of N$_2$ strongly quench scintillation light ~\cite{N2contamination} and it couldn't be removed 
by the ICARUS T600 filtering system. 
As a consequence a custom set-up based on a commercial mass spectrometer (Pfeiffer QMG 220) was specifically developed 
to measure nitrogen contamination in Ar. A sample of the T600 gas phase was periodically analyzed and N$_2$ concentration 
was always found below the 1 ppm sensitivity of the instrument according to evidence from the PMT signals.

Beside LAr purity, other cryogenic parameters affecting the LAr-TPC performance  were accurately monitored 
along the whole detector operation. In particular, the internal temperature, directly connected to the electron drift 
velocity, was found stable and uniform to better than 0.25 K (Fig.~\ref{Ttrend}). This was confirmed by the observed 
stability of the internal absolute pressure (Fig.~\ref{Ptrend}) in spite of the previously  mentioned stops and accidents.



\begin{center}
\includegraphics[width=11cm]{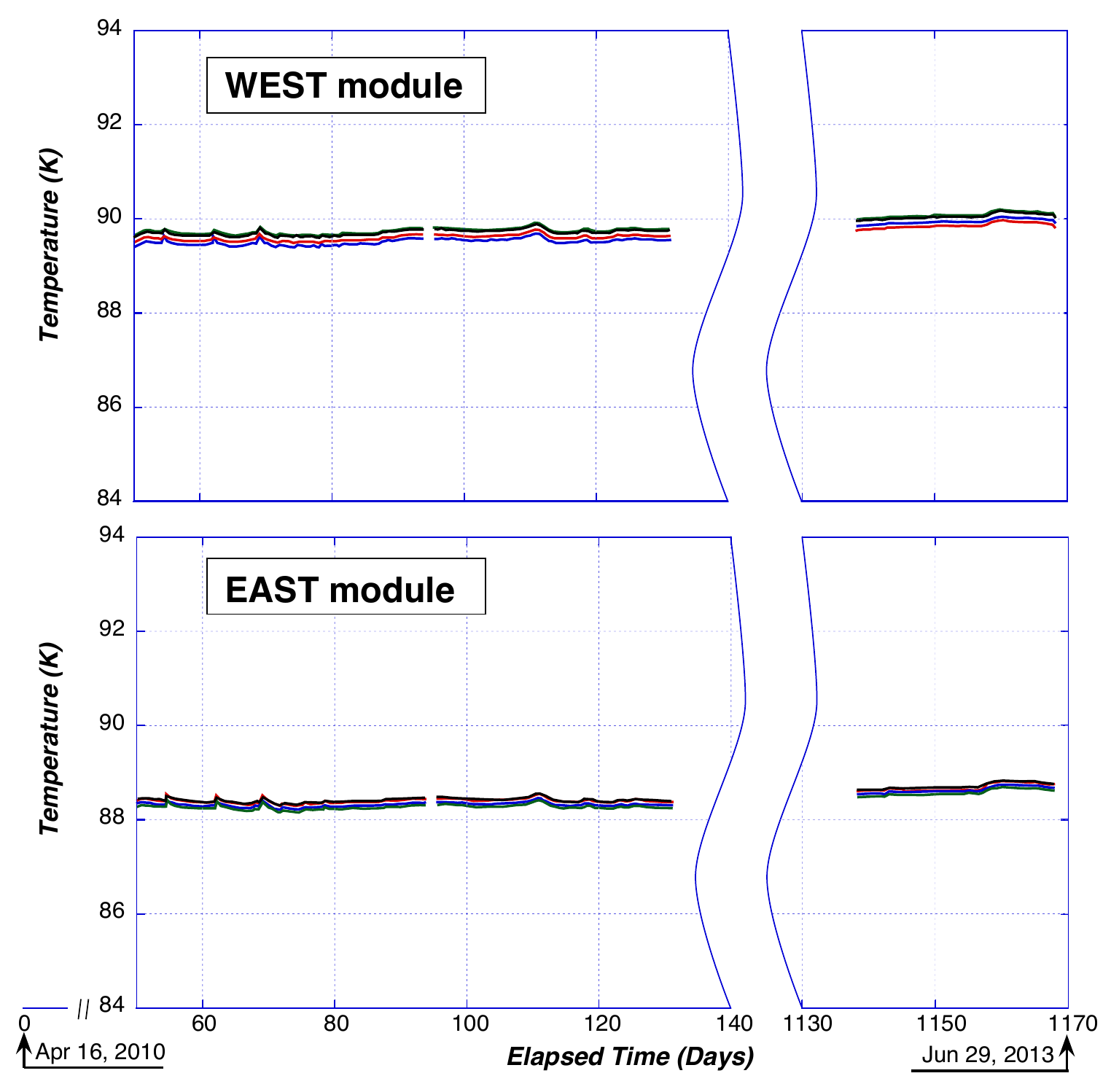}
\captionof{figure}{Trend of the internal temperatures measured in three different vertical positions (bottom, middle height, top) in the 
two modules recorded in two periods of the detector live time, one at the beginning of the run, in 2010, and the second 
in 2013, close to the end of the run.}
\label{Ttrend}
\end{center}

\begin{center}
\includegraphics[width=12cm]{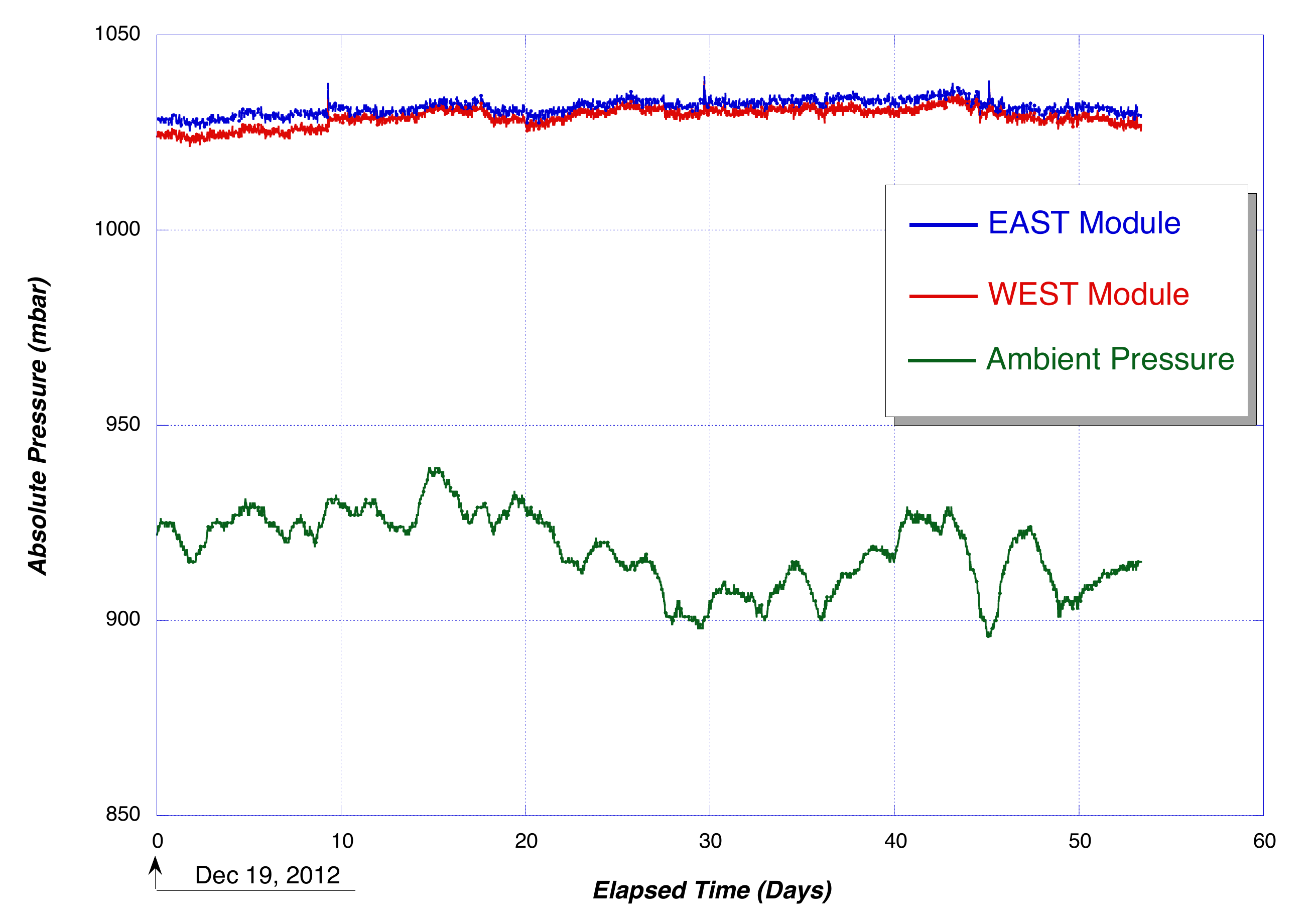}
\captionof{figure}{Absolute internal pressure in the two modules during a period of about two months between the end of 2012 and the 
beginning of 2013. The internal pressure was kept uniform and stable within $\approx$10 mbar.}
\label{Ptrend}
\end{center}

\section{Decommissioning}

\begin{figure}[htbp!]
\begin{center}
\includegraphics[width=14cm]{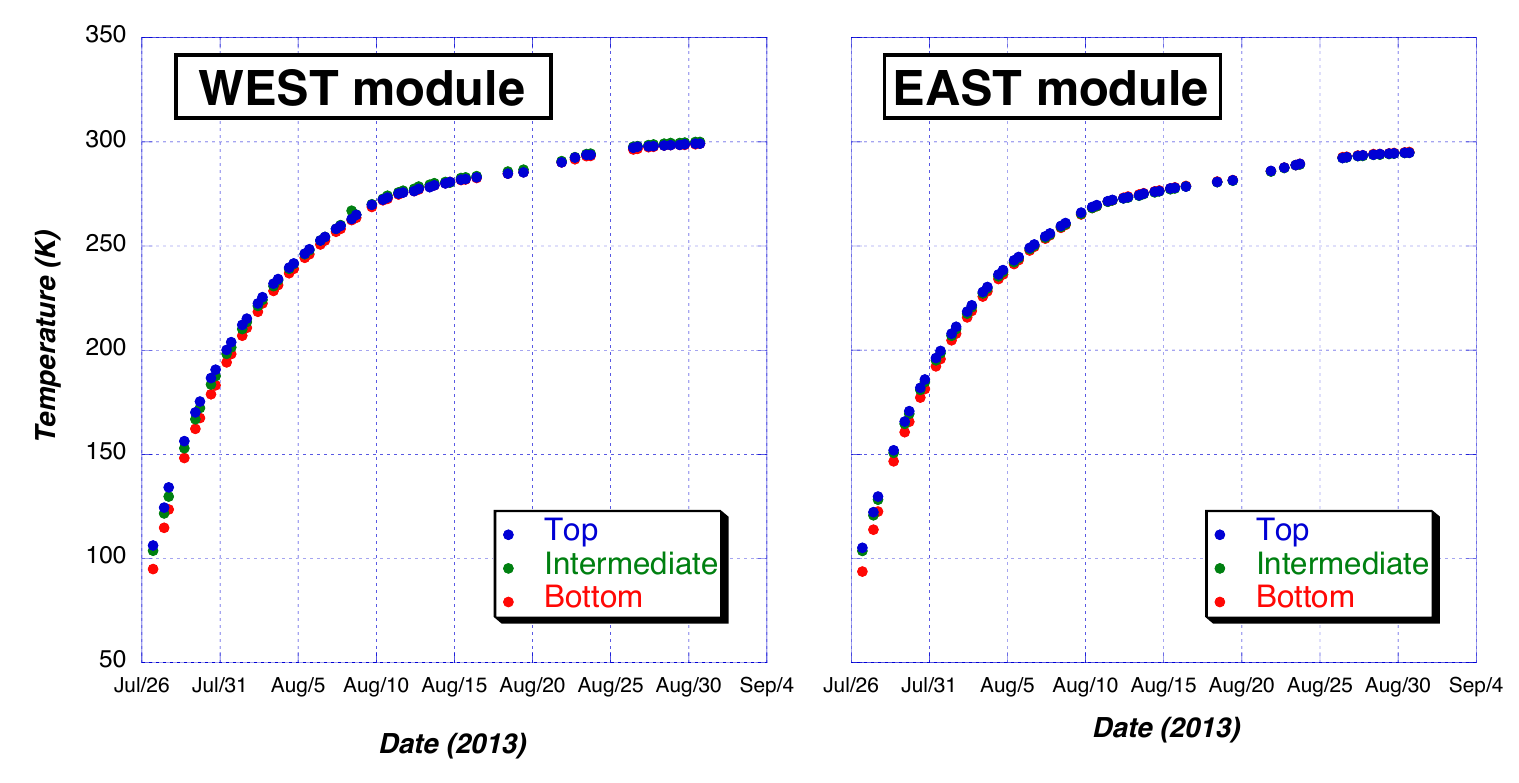}
\caption{Temperature trends on the wire-chamber structure all over the T600 warm-up phase.}
\label{Twarmup}
\end{center}
\end{figure}

The  preparation for the T600 decommissioning was started during the last months of detector operation. In particular an emptying skid was installed to host an immersed LAr pump, together with an intermediate buffer to speed up the emptying process.
The decommissioning process started on June 27$^{th}$ 2013 proceeding with the following phases:

\begin{itemize}

\item[1.] The cryostat emptying phase lasted less than one month and was operated in a safe 
     and smooth way in parallel on the two modules. To speed up the process, LAr was transferred  at 
     7,000 l/h rate into an intermediate vessel (20,000 l), allowing the decoupling of  the emptying 
     procedure from the truck uploading. About 740 LAr tons over a total of 760 were recovered for a total 
     number of 33 trucks (1-2 truck /day). 
\item[2.] The successive cryostat warming-up started on July 25$^{th}$ and took about one month proceeding 
     with the help of a heating system to speed-up the process circulating warm nitrogen gas inside T600 cooling 
     screens while keeping the thermal gradients within the $\Delta$T$_{max}$ < 50 K specification to prevent 
     thermal shock on wire chambers (Fig.~\ref{Twarmup}).
\item[3.] The T600 detector dismantling started in September 2013 and lasted about 15 months.
     It was finalized to the cryostat opening to extract the TPC detectors as a whole including the 
     light detection system, cabling and ancillary equipments, placed into dedicated boxes specifically 
     designed for the transport to CERN. The cryogenic plant, DAQ and trigger electronic systems were 
     recuperated and separately sent to CERN.
\end{itemize}

\section{Conclusions}
\label{conclusions}

The ICARUS T600 LAr-TPC, installed at LNGS, is the biggest liquid argon detector ever realized and represents so far 
the state of the art of the liquid argon TPC technology. Industry partnership was crucial to perform a scaling-up of 
the technology from the laboratory prototypal scale to the kt mass scale. 

The commissioning at LNGS was successfully and safely performed during the first half of 2010.  The detector smoothly 
reached optimal working conditions and took cosmic and CNGS neutrino beam data with extremely high liquid argon purity 
and high detector live-time, performing even beyond expectations. 
The obtained results demonstrated, as reported in several published papers, the effectiveness of the single phase LAr-TPC 
detectors paving the way to the construction 
of huge detectors with longer drift distances.

The three years safe and stable operation in the severe underground environment condition was an important achievement 
for LAr-TPC technique demonstrating  the technology is mature and  scalable to several kton mass as required by future projects. 
 Lessons learned from the plant operation, accidental events and plant improvements will be useful for future developments.

\section{Acknowledgements}
The ICARUS Collaboration acknowledges the fundamental support of INFN and, in particular, of the LNGS Laboratory, all the staff and its Directors, to the construction and operation of the experiment. Moreover the authors thank LNGS Safety and Prevention Service, the Research and Technical Divisions, and in particular the Experiment Support Service, the LNGS cryogenic group and the Exercise and Maintenance Service for their contribution to the commissioning and operation of the T600 apparatus.
The collaboration recognizes the fundamental involvement of the industrial companies Air Liquide, Stirling Cryogenics BV and Luca Scarcia, which contributed in the realization, operation and maintenance of the cryogenic plant.
A special thank to Marco Brugnolli, Arnaldo Di Cesare and Donatello Ciccotti.
The authors warmly thank the Electronics Service of INFN Pavia, in particular M.C. Prata, for the design and realization of the slow control electronics and of the vacuum system remote controls.
The Polish groups acknowledge the support of the National Science Center, Harmonia (2012/04/M/ST2/00775) and Preludium (2011/03/N/ST2/01971) funding schemes.

\end{document}